\documentclass[11pt,twoside]{atmp}

\usepackage{amsmath,amssymb}
\usepackage[all]{xy}
\usepackage{color}

\copyrightnotice{2008}{1}{1}{1} 

\usepackage{amsmath,amsfonts,amsthm,graphicx,maple2e}

\usepackage{amssymb}
\usepackage{amsmath,amsfonts}

\def\flecha{\longrightarrow}
\def\implies{\Longrightarrow}

\def\Tr{\operatorname{Tr}}

\def\OO{\mathcal{O}}
\def\MM{\mathcal{M}}

\def\N{\mathcal{N}}
\def\M{\mathcal{M}}

\def\CC{\mathbb{C}}

\def\PP{\mathbb{P}}

\def\AA{\mathbb{A}}
\def\ZZ{\mathbb{Z}}

\def\for{\mathrm{for}}
\def\and{\mathrm{and}}

\def\where{\mathrm{where}}

\def\even{\mathrm{even}}
\def\odd{\mathrm{odd}}

\def\skp{\\[.2in]}
\def\skop{\;\;\;\;\;\;\;\;}

\def\til{\widetilde}
\def\Hat{\widehat}

\def\part#1{\dfrac{\partial}{\partial #1}}
\def\d{\mathrm{d}}

\def\12{\dfrac{1}{2}}

\def\C{\mathcal{C}}
\def\S{\mathcal{S}}

\makeatletter

\def\diagram{\leftwidth=\z@ \rightwidth=\z@ \topheight=\z@
\botheight=\z@ \setbox\@picbox\hbox\bgroup}

\def\enddiagram{\egroup\wd\@picbox\rightwidth\unitlength
\ht\@picbox\topheight\unitlength \dp\@picbox\botheight\unitlength
\hskip\leftwidth\unitlength\box\@picbox}

\def\bfig{\begin{diagram}}
\def\efig{\end{diagram}}
\newcount\wideness \newcount\leftwidth \newcount\rightwidth
\newcount\highness \newcount\topheight \newcount\botheight

\def\ratchet#1#2{\ifnum#1<#2 \global #1=#2 \fi}

\def\putbox(#1,#2)#3{%
\horsize{\wideness}{#3} \divide\wideness by 2
{\advance\wideness by #1 \ratchet{\rightwidth}{\wideness}}
{\advance\wideness by -#1 \ratchet{\leftwidth}{\wideness}}
\vertsize{\highness}{#3} \divide\highness by 2
{\advance\highness by #2 \ratchet{\topheight}{\highness}}
{\advance\highness by -#2 \ratchet{\botheight}{\highness}}
\put(#1,#2){\makebox(0,0){$#3$}}}

\def\putlbox(#1,#2)#3{%
\horsize{\wideness}{#3}
{\advance\wideness by #1 \ratchet{\rightwidth}{\wideness}}
{\ratchet{\leftwidth}{-#1}}
\vertsize{\highness}{#3} \divide\highness by 2
{\advance\highness by #2 \ratchet{\topheight}{\highness}}
{\advance\highness by -#2 \ratchet{\botheight}{\highness}}
\put(#1,#2){\makebox(0,0)[l]{$#3$}}}

\def\putrbox(#1,#2)#3{%
\horsize{\wideness}{#3}
{\ratchet{\rightwidth}{#1}}
{\advance\wideness by -#1 \ratchet{\leftwidth}{\wideness}}
\vertsize{\highness}{#3} \divide\highness by 2
{\advance\highness by #2 \ratchet{\topheight}{\highness}}
{\advance\highness by -#2 \ratchet{\botheight}{\highness}}
\put(#1,#2){\makebox(0,0)[r]{$#3$}}}

\def\adjust[#1]{} 

\newcount \coefa
\newcount \coefb
\newcount \coefc
\newcount\tempcounta
\newcount\tempcountb
\newcount\tempcountc
\newcount\tempcountd
\newcount\xext
\newcount\yext
\newcount\xoff
\newcount\yoff
\newcount\gap%
\newcount\arrowtypea
\newcount\arrowtypeb
\newcount\arrowtypec
\newcount\arrowtyped
\newcount\arrowtypee
\newcount\height
\newcount\width
\newcount\xpos
\newcount\ypos
\newcount\run
\newcount\rise
\newcount\arrowlength
\newcount\halflength
\newcount\arrowtype
\newdimen\tempdimen
\newdimen\xlen
\newdimen\ylen
\newsavebox{\tempboxa}%
\newsavebox{\tempboxb}%
\newsavebox{\tempboxc}%

\newdimen\w@dth

\def\setw@dth#1#2{\setbox\z@\hbox{$#1$}\w@dth=\wd\z@
\setbox\@ne\hbox{$#2$}\ifnum\w@dth<\wd\@ne \w@dth=\wd\@ne \fi
\advance\w@dth by 1.2em}

\def\t@^#1_#2{\def\n@one{#1}\def\n@two{#2}\mathrel{\setw@dth{#1}{#2}
\mathop{\hbox to \w@dth{\rightarrowfill}}\limits
\ifx\n@one\empty\else ^{\box\z@}\fi
\ifx\n@two\empty\else _{\box\@ne}\fi}}
\def\t@@^#1{\@ifnextchar_ {\t@^{#1}}{\t@^{#1}_{}}}
\def\to{\@ifnextchar^ {\t@@}{\t@@^{}}}

\def\t@left^#1_#2{\def\n@one{#1}\def\n@two{#2}\mathrel{\setw@dth{#1}{#2}
\mathop{\hbox to \w@dth{\leftarrowfill}}\limits
\ifx\n@one\empty\else ^{\box\z@}\fi
\ifx\n@two\empty\else _{\box\@ne}\fi}}
\def\t@@left^#1{\@ifnextchar_ {\t@left^{#1}}{\t@left^{#1}_{}}}
\def\toleft{\@ifnextchar^ {\t@@left}{\t@@left^{}}}

\def\two@^#1_#2{\def\n@one{#1}\def\n@two{#2}\mathrel{\setw@dth{#1}{#2}
\mathop{\vcenter{\hbox to \w@dth{\rightarrowfill}\kern-1.7ex
         \hbox to \w@dth{\rightarrowfill}}%
       }\limits
\ifx\n@one\empty\else ^{\box\z@}\fi
\ifx\n@two\empty\else _{\box\@ne}\fi}}
\def\tw@@^#1{\@ifnextchar_ {\two@^{#1}}{\two@^{#1}_{}}}
\def\two{\@ifnextchar^ {\tw@@}{\tw@@^{}}}

\def\tofr@^#1_#2{\def\n@one{#1}\def\n@two{#2}\mathrel{\setw@dth{#1}{#2}
\mathop{\vcenter{\hbox to \w@dth{\rightarrowfill}\kern-1.7ex
         \hbox to \w@dth{\leftarrowfill}}%
       }\limits
\ifx\n@one\empty\else ^{\box\z@}\fi
\ifx\n@two\empty\else _{\box\@ne}\fi}}
\def\t@fr@^#1{\@ifnextchar_ {\tofr@^{#1}}{\tofr@^{#1}_{}}}
\def\tofro{\@ifnextchar^ {\t@fr@}{\t@fr@^{}}}

\def\mon{\mathop{\m@th\hbox to
      14.6\P@{\lasyb\char'51\hskip-2.1\P@$\arrext$\hss
$\mathord\rightarrow$}}\limits} 
\def\leftmono{\mathrel{\m@th\hbox to
14.6\P@{$\mathord\leftarrow$\hss$\arrext$\hskip-2.1\P@\lasyb\char'50%
}}\limits} 
\mathchardef\arrext="0200       

\def\settypes(#1,#2,#3){\arrowtypea#1 \arrowtypeb#2 \arrowtypec#3}
\def\settoheight#1#2{\setbox\@tempboxa\hbox{#2}#1\ht\@tempboxa\relax}%
\def\settodepth#1#2{\setbox\@tempboxa\hbox{#2}#1\dp\@tempboxa\relax}%
\def\settokens[#1`#2`#3`#4]{%
     \def\tokena{#1}\def\tokenb{#2}\def\tokenc{#3}\def\tokend{#4}}
\def\setsqparms[#1`#2`#3`#4;#5`#6]{%
\arrowtypea #1
\arrowtypeb #2
\arrowtypec #3
\arrowtyped #4
\width #5
\height #6
}
\def\setpos(#1,#2){\xpos=#1 \ypos#2}

\def\settriparms[#1`#2`#3;#4]{\settripairparms[#1`#2`#3`1`1;#4]}%

\def\settripairparms[#1`#2`#3`#4`#5;#6]{%
\arrowtypea #1
\arrowtypeb #2
\arrowtypec #3
\arrowtyped #4
\arrowtypee #5
\width #6
\height #6
}

\def\resetparms{\settripairparms[1`1`1`1`1;500]\width 500}

\resetparms

\def\mvector(#1,#2)#3{
\put(0,0){\vector(#1,#2){#3}}%
\put(0,0){\vector(#1,#2){26}}%
}
\def\evector(#1,#2)#3{{
\arrowlength #3
\put(0,0){\vector(#1,#2){\arrowlength}}%
\advance \arrowlength by-30
\put(0,0){\vector(#1,#2){\arrowlength}}%
}}

\def\horsize#1#2{%
\settowidth{\tempdimen}{$#2$}%
#1=\tempdimen
\divide #1 by\unitlength
}

\def\vertsize#1#2{%
\settoheight{\tempdimen}{$#2$}%
#1=\tempdimen
\settodepth{\tempdimen}{$#2$}%
\advance #1 by\tempdimen
\divide #1 by\unitlength
}

\def\putvector(#1,#2)(#3,#4)#5#6{{%
\ifnum3<\arrowtype
\putdashvector(#1,#2)(#3,#4)#5\arrowtype
\else
\ifnum\arrowtype<-3
\putdashvector(#1,#2)(#3,#4)#5\arrowtype
\else
\xpos=#1
\ypos=#2
\run=#3
\rise=#4
\arrowlength=#5
\ifnum \arrowtype<0
    \ifnum \run=0
    \advance \ypos by-\arrowlength
    \else
    \tempcounta \arrowlength
    \multiply \tempcounta by\rise
    \divide \tempcounta by\run
    \ifnum\run>0
        \advance \xpos by\arrowlength
        \advance \ypos by\tempcounta
    \else
        \advance \xpos by-\arrowlength
        \advance \ypos by-\tempcounta
    \fi
    \fi
    \multiply \arrowtype by-1
    \multiply \rise by-1
    \multiply \run by-1
\fi
\ifcase \arrowtype
\or \put(\xpos,\ypos){\vector(\run,\rise){\arrowlength}}%
\or \put(\xpos,\ypos){\mvector(\run,\rise)\arrowlength}%
\or \put(\xpos,\ypos){\evector(\run,\rise){\arrowlength}}%
\fi\fi\fi
}}

\def\putsplitvector(#1,#2)#3#4{
\xpos #1
\ypos #2
\arrowtype #4
\halflength #3
\arrowlength #3
\gap 140
\advance \halflength by-\gap
\divide \halflength by2
\ifnum\arrowtype>0
   \ifcase \arrowtype
   \or \put(\xpos,\ypos){\line(0,-1){\halflength}}%
       \advance\ypos by-\halflength
       \advance\ypos by-\gap
       \put(\xpos,\ypos){\vector(0,-1){\halflength}}%
   \or \put(\xpos,\ypos){\line(0,-1)\halflength}%
       \put(\xpos,\ypos){\vector(0,-1)3}%
       \advance\ypos by-\halflength
       \advance\ypos by-\gap
       \put(\xpos,\ypos){\vector(0,-1){\halflength}}%
   \or \put(\xpos,\ypos){\line(0,-1)\halflength}%
       \advance\ypos by-\halflength
       \advance\ypos by-\gap
       \put(\xpos,\ypos){\evector(0,-1){\halflength}}%
   \fi
\else \arrowtype=-\arrowtype
   \ifcase\arrowtype
   \or \advance \ypos by-\arrowlength
       \put(\xpos,\ypos){\line(0,1){\halflength}}%
       \advance\ypos by\halflength
       \advance\ypos by\gap
       \put(\xpos,\ypos){\vector(0,1){\halflength}}%
   \or \advance \ypos by-\arrowlength
       \put(\xpos,\ypos){\line(0,1)\halflength}%
       \put(\xpos,\ypos){\vector(0,1)3}%
       \advance\ypos by\halflength
       \advance\ypos by\gap
       \put(\xpos,\ypos){\vector(0,1){\halflength}}%
   \or \advance \ypos by-\arrowlength
       \put(\xpos,\ypos){\line(0,1)\halflength}%
       \advance\ypos by\halflength
       \advance\ypos by\gap
       \put(\xpos,\ypos){\evector(0,1){\halflength}}%
   \fi
\fi
}

\def\putmorphism(#1)(#2,#3)[#4`#5`#6]#7#8#9{{%
\run #2
\rise #3
\ifnum\rise=0
  \puthmorphism(#1)[#4`#5`#6]{#7}{#8}#9%
\else\ifnum\run=0
  \putvmorphism(#1)[#4`#5`#6]{#7}{#8}#9%
\else
\setpos(#1)%
\arrowlength #7
\arrowtype #8
\ifnum\run=0
\else\ifnum\rise=0
\else
\ifnum\run>0
    \coefa=1
\else
   \coefa=-1
\fi
\ifnum\arrowtype>0
   \coefb=0
   \coefc=-1
\else
   \coefb=\coefa
   \coefc=1
   \arrowtype=-\arrowtype
\fi
\width=2
\multiply \width by\run
\divide \width by\rise
\ifnum \width<0  \width=-\width\fi
\advance\width by60
\if l#9 \width=-\width\fi
\putbox(\xpos,\ypos){#4}
{\multiply \coefa by\arrowlength
\advance\xpos by\coefa
\multiply \coefa by\rise
\divide \coefa by\run
\advance \ypos by\coefa
\putbox(\xpos,\ypos){#5} }%
{\multiply \coefa by\arrowlength
\divide \coefa by2
\advance \xpos by\coefa
\advance \xpos by\width
\multiply \coefa by\rise
\divide \coefa by\run
\advance \ypos by\coefa
\if l#9%
   \putrbox(\xpos,\ypos){#6}%
\else\if r#9%
   \putlbox(\xpos,\ypos){#6}%
\fi\fi }%
{\multiply \rise by-\coefc
\multiply \run by-\coefc
\multiply \coefb by\arrowlength
\advance \xpos by\coefb
\multiply \coefb by\rise
\divide \coefb by\run
\advance \ypos by\coefb
\multiply \coefc by70
\advance \ypos by\coefc
\multiply \coefc by\run
\divide \coefc by\rise
\advance \xpos by\coefc
\multiply \coefa by140
\multiply \coefa by\run
\divide \coefa by\rise
\advance \arrowlength by\coefa
\ifcase\arrowtype
\or \put(\xpos,\ypos){\vector(\run,\rise){\arrowlength}}%
\or \put(\xpos,\ypos){\mvector(\run,\rise){\arrowlength}}%
\or \put(\xpos,\ypos){\evector(\run,\rise){\arrowlength}}%
\fi}\fi\fi\fi\fi}}

\newcount\numbdashes \newcount\lengthdash \newcount\increment

\def\howmanydashes{
\numbdashes=\arrowlength \lengthdash=40
\divide\numbdashes by \lengthdash
\lengthdash=\arrowlength
\divide\lengthdash by \numbdashes
\increment=\lengthdash
\multiply\lengthdash by 3
\divide\lengthdash by 5
}

\def\putdashvector(#1)(#2,#3)#4#5{%
\ifnum#3=0 \putdashhvector(#1){#4}#5
\else
\ifnum#2=0
\putdashvvector(#1){#4}#5\fi\fi}

\def\putdashhvector(#1,#2)#3#4{{%
\arrowlength=#3 \howmanydashes
\multiput(#1,#2)(\increment,0){\numbdashes}%
{\vrule height .4pt width \lengthdash\unitlength}
\arrowtype=#4 \xpos=#1
\ifnum\arrowtype<0 \advance\arrowtype by 7 \fi
\ifcase\arrowtype
\or \advance\xpos by 10
    \put(\xpos,#2){\vector(-1,0){\lengthdash}}
    \advance\xpos by 40
    \put(\xpos,#2){\vector(-1,0){\lengthdash}}
\or \advance \xpos by 10
    \put(\xpos,#2){\vector(-1,0){\lengthdash}}
    \advance\xpos by  \arrowlength
    \advance\xpos by  -50
    \put(\xpos,#2){\vector(-1,0){\lengthdash}}
\or \advance\xpos by 10
    \put(\xpos,#2){\vector(-1,0){\lengthdash}}
\or \advance\xpos by \arrowlength
    \advance\xpos by -\lengthdash
    \put(\xpos,#2){\vector(1,0){\lengthdash}}
\or {\advance\xpos by 10
    \put(\xpos,#2){\vector(1,0){\lengthdash}}}
    \advance\xpos by \arrowlength
    \advance\xpos by -\lengthdash
    \put(\xpos,#2){\vector(1,0){\lengthdash}}
\or \advance\xpos by \arrowlength
    \advance\xpos by -\lengthdash
    \put(\xpos,#2){\vector(1,0){\lengthdash}}
    \advance\xpos by -40
    \put(\xpos,#2){\vector(1,0){\lengthdash}}
   \fi
}}

\def\putdashvvector(#1,#2)#3#4{{%
\arrowlength=#3 \howmanydashes
\ypos=#2 \advance\ypos by -\arrowlength
\multiput(#1,#2)(0,\increment){\numbdashes}%
    {\vrule width .4pt height \lengthdash\unitlength}
\arrowtype=#4 \ypos=#2
\ifnum\arrowtype<0 \advance\arrowtype by 7 \fi
\ifcase\arrowtype
\or \advance\ypos by \arrowlength \advance\ypos by -40
    \put(#1,\ypos){\vector(0,1){\lengthdash}}
    \advance\ypos by -40
    \put(#1,\ypos){\vector(0,1){\lengthdash}}
\or \advance\ypos by 10
    \put(#1,\ypos){\vector(0,1){\lengthdash}}
    \advance\ypos by \arrowlength \advance\ypos by -40
    \put(#1,\ypos){\vector(0,1){\lengthdash}}
\or \advance\ypos by \arrowlength \advance\ypos by -40
    \put(#1,\ypos){\vector(0,1){\lengthdash}}
\or \advance\ypos by 10
    \put(#1,\ypos){\vector(0,-1){\lengthdash}}
\or \advance\ypos by 10
    \put(#1,\ypos){\vector(0,-1){\lengthdash}}
    \advance\ypos by \arrowlength \advance\ypos by -40
    \put(#1,\ypos){\vector(0,-1){\lengthdash}}
\or \advance\ypos by 10
    \put(#1,\ypos){\vector(0,-1){\lengthdash}}
    \advance\ypos by 40
    \put(#1,\ypos){\vector(0,-1){\lengthdash}}
\fi
}}

\def\puthmorphism(#1,#2)[#3`#4`#5]#6#7#8{{%
\xpos #1
\ypos #2
\width #6
\arrowlength #6
\arrowtype=#7
\putbox(\xpos,\ypos){#3\vphantom{#4}}%
{\advance \xpos by\arrowlength
\putbox(\xpos,\ypos){\vphantom{#3}#4}}%
\horsize{\tempcounta}{#3}%
\horsize{\tempcountb}{#4}%
\divide \tempcounta by2
\divide \tempcountb by2
\advance \tempcounta by30
\advance \tempcountb by30
\advance \xpos by\tempcounta
\advance \arrowlength by-\tempcounta
\advance \arrowlength by-\tempcountb
\putvector(\xpos,\ypos)(1,0)\arrowlength\arrowtype
\divide \arrowlength by2
\advance \xpos by\arrowlength
\vertsize{\tempcounta}{#5}%
\divide\tempcounta by2
\advance \tempcounta by20
\if a#8 %
   \advance \ypos by\tempcounta
   \putbox(\xpos,\ypos){#5}%
\else
   \advance \ypos by-\tempcounta
   \putbox(\xpos,\ypos){#5}%
\fi}}

\def\putvmorphism(#1,#2)[#3`#4`#5]#6#7#8{{%
\xpos #1
\ypos #2
\arrowlength #6
\arrowtype #7
\settowidth{\xlen}{$#5$}%
\putbox(\xpos,\ypos){#3}%
{\advance \ypos by-\arrowlength
\putbox(\xpos,\ypos){#4}}%
{\advance\arrowlength by-140
\advance \ypos by-70
\ifdim\xlen>0pt
   \if m#8%
      \putsplitvector(\xpos,\ypos)\arrowlength\arrowtype
   \else
   \putvector(\xpos,\ypos)(0,-1)\arrowlength\arrowtype
   \fi
\else
   \putvector(\xpos,\ypos)(0,-1)\arrowlength\arrowtype
\fi}%
\ifdim\xlen>0pt
   \divide \arrowlength by2
   \advance\ypos by-\arrowlength
   \if l#8%
      \advance \xpos by-40
      \putrbox(\xpos,\ypos){#5}%
   \else\if r#8%
      \advance \xpos by40
      \putlbox(\xpos,\ypos){#5}%
   \else
      \putbox(\xpos,\ypos){#5}%
   \fi\fi
\fi
}}

\def\putsquarep<#1>(#2)[#3;#4`#5`#6`#7]{{%
\setsqparms[#1]%
\setpos(#2)%
\settokens[#3]%
\puthmorphism(\xpos,\ypos)[\tokenc`\tokend`{#7}]{\width}{\arrowtyped}b%
\advance\ypos by \height
\puthmorphism(\xpos,\ypos)[\tokena`\tokenb`{#4}]{\width}{\arrowtypea}a%
\putvmorphism(\xpos,\ypos)[``{#5}]{\height}{\arrowtypeb}l%
\advance\xpos by \width
\putvmorphism(\xpos,\ypos)[``{#6}]{\height}{\arrowtypec}r%
}}

\def\putsquare{\@ifnextchar <{\putsquarep}{\putsquarep%
   <\arrowtypea`\arrowtypeb`\arrowtypec`\arrowtyped;\width`\height>}}
\def\square{\@ifnextchar< {\squarep}{\squarep
   <\arrowtypea`\arrowtypeb`\arrowtypec`\arrowtyped;\width`\height>}}
\def\squarep<#1>[#2`#3`#4`#5;#6`#7`#8`#9]{{
\setsqparms[#1]
\diagram
\putsquarep<\arrowtypea`\arrowtypeb`\arrowtypec`
\arrowtyped;\width`\height>
(0,0)[#2`#3`#4`{#5};#6`#7`#8`{#9}]
\enddiagram
}}                                                 
\def\putptrianglep<#1>(#2,#3)[#4`#5`#6;#7`#8`#9]{{%
\settriparms[#1]%
\xpos=#2 \ypos=#3
\advance\ypos by \height
\puthmorphism(\xpos,\ypos)[#4`#5`{#7}]{\height}{\arrowtypea}a%
\putvmorphism(\xpos,\ypos)[`#6`{#8}]{\height}{\arrowtypeb}l%
\advance\xpos by\height
\putmorphism(\xpos,\ypos)(-1,-1)[``{#9}]{\height}{\arrowtypec}r%
}}

\def\putptriangle{\@ifnextchar <{\putptrianglep}{\putptrianglep
   <\arrowtypea`\arrowtypeb`\arrowtypec;\height>}}
\def\ptriangle{\@ifnextchar <{\ptrianglep}{\ptrianglep
   <\arrowtypea`\arrowtypeb`\arrowtypec;\height>}}
\def\ptrianglep<#1>[#2`#3`#4;#5`#6`#7]{{
\settriparms[#1]
\diagram
\putptrianglep<\arrowtypea`\arrowtypeb`
\arrowtypec;\height>
(0,0)[#2`#3`#4;#5`#6`{#7}]
\enddiagram
}}                                            

\def\putqtrianglep<#1>(#2,#3)[#4`#5`#6;#7`#8`#9]{{%
\settriparms[#1]%
\xpos=#2 \ypos=#3
\advance\ypos by\height
\puthmorphism(\xpos,\ypos)[#4`#5`{#7}]{\height}{\arrowtypea}a%
\putmorphism(\xpos,\ypos)(1,-1)[``{#8}]{\height}{\arrowtypeb}l%
\advance\xpos by\height
\putvmorphism(\xpos,\ypos)[`#6`{#9}]{\height}{\arrowtypec}r%
}}

\def\putqtriangle{\@ifnextchar <{\putqtrianglep}{\putqtrianglep
   <\arrowtypea`\arrowtypeb`\arrowtypec;\height>}}
\def\qtriangle{\@ifnextchar <{\qtrianglep}{\qtrianglep
   <\arrowtypea`\arrowtypeb`\arrowtypec;\height>}}
\def\qtrianglep<#1>[#2`#3`#4;#5`#6`#7]{{
\settriparms[#1]
\width=\height                                
\diagram
\putqtrianglep<\arrowtypea`\arrowtypeb`
\arrowtypec;\height>
(0,0)[#2`#3`#4;#5`#6`{#7}]
\enddiagram
}}

\def\putdtrianglep<#1>(#2,#3)[#4`#5`#6;#7`#8`#9]{{%
\settriparms[#1]%
\xpos=#2 \ypos=#3
\puthmorphism(\xpos,\ypos)[#5`#6`{#9}]{\height}{\arrowtypec}b%
\advance\xpos by \height \advance\ypos by\height
\putmorphism(\xpos,\ypos)(-1,-1)[``{#7}]{\height}{\arrowtypea}l%
\putvmorphism(\xpos,\ypos)[#4``{#8}]{\height}{\arrowtypeb}r%
}}

\def\putdtriangle{\@ifnextchar <{\putdtrianglep}{\putdtrianglep
   <\arrowtypea`\arrowtypeb`\arrowtypec;\height>}}
\def\dtriangle{\@ifnextchar <{\dtrianglep}{\dtrianglep
   <\arrowtypea`\arrowtypeb`\arrowtypec;\height>}}
\def\dtrianglep<#1>[#2`#3`#4;#5`#6`#7]{{
\settriparms[#1]
\width=\height                                
\diagram
\putdtrianglep<\arrowtypea`\arrowtypeb`
\arrowtypec;\height>
(0,0)[#2`#3`#4;#5`#6`{#7}]
\enddiagram
}}

\def\putbtrianglep<#1>(#2,#3)[#4`#5`#6;#7`#8`#9]{{%
\settriparms[#1]%
\xpos=#2 \ypos=#3
\puthmorphism(\xpos,\ypos)[#5`#6`{#9}]{\height}{\arrowtypec}b%
\advance\ypos by\height
\putmorphism(\xpos,\ypos)(1,-1)[``{#8}]{\height}{\arrowtypeb}r%
\putvmorphism(\xpos,\ypos)[#4``{#7}]{\height}{\arrowtypea}l%
}}

\def\putbtriangle{\@ifnextchar <{\putbtrianglep}{\putbtrianglep
   <\arrowtypea`\arrowtypeb`\arrowtypec;\height>}}
\def\btriangle{\@ifnextchar <{\btrianglep}{\btrianglep
   <\arrowtypea`\arrowtypeb`\arrowtypec;\height>}}
\def\btrianglep<#1>[#2`#3`#4;#5`#6`#7]{{
\settriparms[#1]
\width=\height                               
\diagram
\putbtrianglep<\arrowtypea`\arrowtypeb`
\arrowtypec;\height>
(0,0)[#2`#3`#4;#5`#6`{#7}]
\enddiagram
}}

\def\putAtrianglep<#1>(#2,#3)[#4`#5`#6;#7`#8`#9]{{%
\settriparms[#1]%
\xpos=#2 \ypos=#3
{\multiply \height by2
\puthmorphism(\xpos,\ypos)[#5`#6`{#9}]{\height}{\arrowtypec}b}%
\advance\xpos by\height \advance\ypos by\height
\putmorphism(\xpos,\ypos)(-1,-1)[#4``{#7}]{\height}{\arrowtypea}l%
\putmorphism(\xpos,\ypos)(1,-1)[``{#8}]{\height}{\arrowtypeb}r%
}}

\def\putAtriangle{\@ifnextchar <{\putAtrianglep}{\putAtrianglep
   <\arrowtypea`\arrowtypeb`\arrowtypec;\height>}}
\def\Atriangle{\@ifnextchar <{\Atrianglep}{\Atrianglep
   <\arrowtypea`\arrowtypeb`\arrowtypec;\height>}}
\def\Atrianglep<#1>[#2`#3`#4;#5`#6`#7]{{
\settriparms[#1]
\width=\height                                     
\diagram
\putAtrianglep<\arrowtypea`\arrowtypeb`
\arrowtypec;\height>
(0,0)[#2`#3`#4;#5`#6`{#7}]
\enddiagram
}}

\def\putAtrianglepairp<#1>(#2)[#3;#4`#5`#6`#7`#8]{{%
\settripairparms[#1]%
\setpos(#2)%
\settokens[#3]%
\puthmorphism(\xpos,\ypos)[\tokenb`\tokenc`{#7}]{\height}{\arrowtyped}b%
\advance\xpos by\height
\puthmorphism(\xpos,\ypos)[\phantom{\tokenc}`\tokend`{#8}]%
{\height}{\arrowtypee}b%
\advance\ypos by\height
\putmorphism(\xpos,\ypos)(-1,-1)[\tokena``{#4}]{\height}{\arrowtypea}l%
\putvmorphism(\xpos,\ypos)[``{#5}]{\height}{\arrowtypeb}m%
\putmorphism(\xpos,\ypos)(1,-1)[``{#6}]{\height}{\arrowtypec}r%
}}

\def\putAtrianglepair{\@ifnextchar <{\putAtrianglepairp}{\putAtrianglepairp%
   <\arrowtypea`\arrowtypeb`\arrowtypec`\arrowtyped`\arrowtypee;\height>}}
\def\Atrianglepair{\@ifnextchar <{\Atrianglepairp}{\Atrianglepairp%
   <\arrowtypea`\arrowtypeb`\arrowtypec`\arrowtyped`\arrowtypee;\height>}}

\def\Atrianglepairp<#1>[#2;#3`#4`#5`#6`#7]{{
\settripairparms[#1]
\settokens[#2]
\width=\height                                
\diagram
\putAtrianglepairp                            
<\arrowtypea`\arrowtypeb`\arrowtypec`
\arrowtyped`\arrowtypee;\height>
(0,0)[{#2};#3`#4`#5`#6`{#7}]
\enddiagram
}}

\def\putVtrianglep<#1>(#2,#3)[#4`#5`#6;#7`#8`#9]{{%
\settriparms[#1]%
\xpos=#2 \ypos=#3
\advance\ypos by\height
{\multiply\height by2
\puthmorphism(\xpos,\ypos)[#4`#5`{#7}]{\height}{\arrowtypea}a}%
\putmorphism(\xpos,\ypos)(1,-1)[`#6`{#8}]{\height}{\arrowtypeb}l%
\advance\xpos by\height
\advance\xpos by\height
\putmorphism(\xpos,\ypos)(-1,-1)[``{#9}]{\height}{\arrowtypec}r%
}}

\def\putVtriangle{\@ifnextchar <{\putVtrianglep}{\putVtrianglep
   <\arrowtypea`\arrowtypeb`\arrowtypec;\height>}}
\def\Vtriangle{\@ifnextchar <{\Vtrianglep}{\Vtrianglep
   <\arrowtypea`\arrowtypeb`\arrowtypec;\height>}}
\def\Vtrianglep<#1>[#2`#3`#4;#5`#6`#7]{{
\settriparms[#1]
\width=\height                                 
\diagram
\putVtrianglep<\arrowtypea`\arrowtypeb`
\arrowtypec;\height>
(0,0)[#2`#3`#4;#5`#6`{#7}]
\enddiagram
}}

\def\putVtrianglepairp<#1>(#2)[#3;#4`#5`#6`#7`#8]{{
\settripairparms[#1]%
\setpos(#2)%
\settokens[#3]%
\advance\ypos by\height
\putmorphism(\xpos,\ypos)(1,-1)[`\tokend`{#6}]{\height}{\arrowtypec}l%
\puthmorphism(\xpos,\ypos)[\tokena`\tokenb`{#4}]{\height}{\arrowtypea}a%
\advance\xpos by\height
\puthmorphism(\xpos,\ypos)[\phantom{\tokenb}`\tokenc`{#5}]%
{\height}{\arrowtypeb}a%
\putvmorphism(\xpos,\ypos)[``{#7}]{\height}{\arrowtyped}m%
\advance\xpos by\height
\putmorphism(\xpos,\ypos)(-1,-1)[``{#8}]{\height}{\arrowtypee}r%
}}

\def\putVtrianglepair{\@ifnextchar <{\putVtrianglepairp}{\putVtrianglepairp%
    <\arrowtypea`\arrowtypeb`\arrowtypec`\arrowtyped`\arrowtypee;\height>}}
\def\Vtrianglepair{\@ifnextchar <{\Vtrianglepairp}{\Vtrianglepairp%
    <\arrowtypea`\arrowtypeb`\arrowtypec`\arrowtyped`\arrowtypee;\height>}}
\def\Vtrianglepairp<#1>[#2;#3`#4`#5`#6`#7]{{
\settripairparms[#1]
\settokens[#2]
\diagram
\putVtrianglepairp                             
<\arrowtypea`\arrowtypeb`\arrowtypec`
\arrowtyped`\arrowtypee;\height>
(0,0)[{#2};#3`#4`#5`#6`{#7}]
\enddiagram
}}

\def\putCtrianglep<#1>(#2,#3)[#4`#5`#6;#7`#8`#9]{{%
\settriparms[#1]%
\xpos=#2 \ypos=#3
\advance\ypos by\height
\putmorphism(\xpos,\ypos)(1,-1)[``{#9}]{\height}{\arrowtypec}l%
\advance\xpos by\height
\advance\ypos by\height
\putmorphism(\xpos,\ypos)(-1,-1)[#4`#5`{#7}]{\height}{\arrowtypea}l%
{\multiply\height by 2
\putvmorphism(\xpos,\ypos)[`#6`{#8}]{\height}{\arrowtypeb}r}%
}}

\def\putCtriangle{\@ifnextchar <{\putCtrianglep}{\putCtrianglep
    <\arrowtypea`\arrowtypeb`\arrowtypec;\height>}}
\def\Ctriangle{\@ifnextchar <{\Ctrianglep}{\Ctrianglep
    <\arrowtypea`\arrowtypeb`\arrowtypec;\height>}}
\def\Ctrianglep<#1>[#2`#3`#4;#5`#6`#7]{{
\settriparms[#1]
\width=\height                               
\diagram
\putCtrianglep<\arrowtypea`\arrowtypeb`
\arrowtypec;\height>
(0,0)[#2`#3`#4;#5`#6`{#7}]
\enddiagram
}}                                           
\def\putDtrianglep<#1>(#2,#3)[#4`#5`#6;#7`#8`#9]{{%
\settriparms[#1]%
\xpos=#2 \ypos=#3
\advance\xpos by\height \advance\ypos by\height
\putmorphism(\xpos,\ypos)(-1,-1)[``{#9}]{\height}{\arrowtypec}r%
\advance\xpos by-\height \advance\ypos by\height
\putmorphism(\xpos,\ypos)(1,-1)[`#5`{#8}]{\height}{\arrowtypeb}r%
{\multiply\height by 2
\putvmorphism(\xpos,\ypos)[#4`#6`{#7}]{\height}{\arrowtypea}l}%
}}

\def\putDtriangle{\@ifnextchar <{\putDtrianglep}{\putDtrianglep
    <\arrowtypea`\arrowtypeb`\arrowtypec;\height>}}
\def\Dtriangle{\@ifnextchar <{\Dtrianglep}{\Dtrianglep
   <\arrowtypea`\arrowtypeb`\arrowtypec;\height>}}
\def\Dtrianglep<#1>[#2`#3`#4;#5`#6`#7]{{
\settriparms[#1]
\width=\height                              
\diagram
\putDtrianglep<\arrowtypea`\arrowtypeb`
\arrowtypec;\height>
(0,0)[#2`#3`#4;#5`#6`{#7}]
\enddiagram
}}                                          
\def\setrecparms[#1`#2]{\width=#1 \height=#2}%

\def\recursep<#1`#2>[#3;#4`#5`#6`#7`#8]{{%
\width=#1 \height=#2
\settokens[#3]
\settowidth{\tempdimen}{$\tokena$}
\ifdim\tempdimen=0pt
  \savebox{\tempboxa}{\hbox{$\tokenb$}}%
  \savebox{\tempboxb}{\hbox{$\tokend$}}%
  \savebox{\tempboxc}{\hbox{$#6$}}%
\else
  \savebox{\tempboxa}{\hbox{$\hbox{$\tokena$}\times\hbox{$\tokenb$}$}}%
  \savebox{\tempboxb}{\hbox{$\hbox{$\tokena$}\times\hbox{$\tokend$}$}}%
  \savebox{\tempboxc}{\hbox{$\hbox{$\tokena$}\times\hbox{$#6$}$}}%
\fi
\ypos=\height
\divide\ypos by 2
\xpos=\ypos
\advance\xpos by \width
\bfig
\putCtrianglep<-1`1`1;\ypos>(0,0)[`\tokenc`;#5`#6`{#7}]%
\puthmorphism(\ypos,0)[\tokend`\usebox{\tempboxb}`{#8}]{\width}{-1}b%
\puthmorphism(\ypos,\height)[\tokenb`\usebox{\tempboxa}`{#4}]{\width}{-1}a%
\advance\ypos by \width
\putvmorphism(\ypos,\height)[``\usebox{\tempboxc}]{\height}1r%
\efig
}}

\def\recurse{\@ifnextchar <{\recursep}{\recursep<\width`\height>}}

\def\puttwohmorphisms(#1,#2)[#3`#4;#5`#6]#7#8#9{{%
\puthmorphism(#1,#2)[#3`#4`]{#7}0a
\ypos=#2
\advance\ypos by 20
\puthmorphism(#1,\ypos)[\phantom{#3}`\phantom{#4}`#5]{#7}{#8}a
\advance\ypos by -40
\puthmorphism(#1,\ypos)[\phantom{#3}`\phantom{#4}`#6]{#7}{#9}b
}}

\def\puttwovmorphisms(#1,#2)[#3`#4;#5`#6]#7#8#9{{%
\putvmorphism(#1,#2)[#3`#4`]{#7}0a
\xpos=#1
\advance\xpos by -20
\putvmorphism(\xpos,#2)[\phantom{#3}`\phantom{#4}`#5]{#7}{#8}l
\advance\xpos by 40
\putvmorphism(\xpos,#2)[\phantom{#3}`\phantom{#4}`#6]{#7}{#9}r
}}

\def\puthcoequalizer(#1)[#2`#3`#4;#5`#6`#7]#8#9{{%
\setpos(#1)%
\puttwohmorphisms(\xpos,\ypos)[#2`#3;#5`#6]{#8}11%
\advance\xpos by #8
\puthmorphism(\xpos,\ypos)[\phantom{#3}`#4`#7]{#8}1{#9}
}}

\def\putvcoequalizer(#1)[#2`#3`#4;#5`#6`#7]#8#9{{%
\setpos(#1)%
\puttwovmorphisms(\xpos,\ypos)[#2`#3;#5`#6]{#8}11%
\advance\ypos by -#8
\putvmorphism(\xpos,\ypos)[\phantom{#3}`#4`#7]{#8}1{#9}
}}

\def\putthreehmorphisms(#1)[#2`#3;#4`#5`#6]#7(#8)#9{{%
\setpos(#1) \settypes(#8)
\if a#9 %
     \vertsize{\tempcounta}{#5}%
     \vertsize{\tempcountb}{#6}%
     \ifnum \tempcounta<\tempcountb \tempcounta=\tempcountb \fi
\else
     \vertsize{\tempcounta}{#4}%
     \vertsize{\tempcountb}{#5}%
     \ifnum \tempcounta<\tempcountb \tempcounta=\tempcountb \fi
\fi
\advance \tempcounta by 60
\puthmorphism(\xpos,\ypos)[#2`#3`#5]{#7}{\arrowtypeb}{#9}
\advance\ypos by \tempcounta
\puthmorphism(\xpos,\ypos)[\phantom{#2}`\phantom{#3}`#4]{#7}{\arrowtypea}{#9}
\advance\ypos by -\tempcounta \advance\ypos by -\tempcounta
\puthmorphism(\xpos,\ypos)[\phantom{#2}`\phantom{#3}`#6]{#7}{\arrowtypec}{#9}
}}

\def\setarrowtoks[#1`#2`#3`#4`#5`#6]{%
\def\toka{#1}
\def\tokb{#2}
\def\tokc{#3}
\def\tokd{#4}
\def\toke{#5}
\def\tokf{#6}
}
\def\hex{\@ifnextchar <{\hexp}{\hexp<1000`400>}}
\def\hexp<#1`#2>[#3`#4`#5`#6`#7`#8;#9]{%
\setarrowtoks[#9]
\yext=#2 \advance \yext by #2
\xext=#1 \advance\xext by \yext
\bfig
\putCtriangle<-1`0`1;#2>(0,0)[`#5`;\tokb``\tokd]
\xext=#1 \yext=#2 \advance \yext by #2
\putsquare<1`0`0`1;\xext`\yext>(#2,0)[#3`#4`#7`#8;\toka```\tokf]
\advance \xext by #2
\putDtriangle<0`1`-1;#2>(\xext,0)[`#6`;`\tokc`\toke]
\efig
}

\def\skp{}

\newtheorem*{conjecture}{Conjecture}
\newtheorem{myconjecture}{Conjecture}
\newtheorem{myproposition}{Proposition}
\newtheorem{myproposition2}{Proposition}
\newtheorem*{theorem}{Theorem}
\newtheorem{mytheorem}{Theorem}

\newtheorem*{corollary}{Corollary}

\begin{document}

\title[Superpotentials and ADE singularities]
{Matrix model superpotentials and ADE singularities}

\arxurl{hep-th/}

\author[C. Curto]{Carina Curto}

\address{Department of Mathematics, Duke University, Durham, NC\\
CMBN, Rutgers University, Newark, NJ}  
\addressemail{ccurto@post.harvard.edu}

\begin{abstract}
We use F. Ferrari's methods relating matrix models to Calabi-Yau
spaces in order to explain much of Intriligator and
Wecht's ADE classification of $\N=1$ superconformal theories which
arise as RG fixed points of $\N = 1$ SQCD theories with adjoints.  
We find that ADE superpotentials in
the Intriligator--Wecht classification exactly match matrix model
superpotentials obtained from Calabi-Yaus with corresponding ADE
singularities. Moreover, in the additional $\Hat{O}, \Hat{A}, \Hat{D}$ and $\Hat{E}$ cases
we find new singular geometries.  These `hat' geometries are closely related to their ADE counterparts, 
but feature non-isolated singularities. As a byproduct, we give simple descriptions for small resolutions of Gorenstein threefold singularities in terms of transition functions between just two coordinate charts. To obtain these results we develop an algorithm for blowing down exceptional $\PP^1$s, described in the appendix.
\end{abstract}

\maketitle

\vspace*{2.5in}
\cutpage 
\setcounter{page}{2}

\section{Introduction}

\renewcommand{\a}{\alpha}
\renewcommand{\c}{\gamma}
\renewcommand{\d}{\delta}
\newcommand{\e}{\varepsilon}

Duality has long played an important role in string theory.  In addition, by relating physical quantities (correlators, partition functions, spectra) between different theories with geometric input, dualities have uncovered many unexpected patterns in geometry. This has led to surprising conjectures (such as mirror symmetry and T-duality) which not only have important implications for physics, but are interesting and meaningful in a purely geometric light.

\begin{figure}[h]
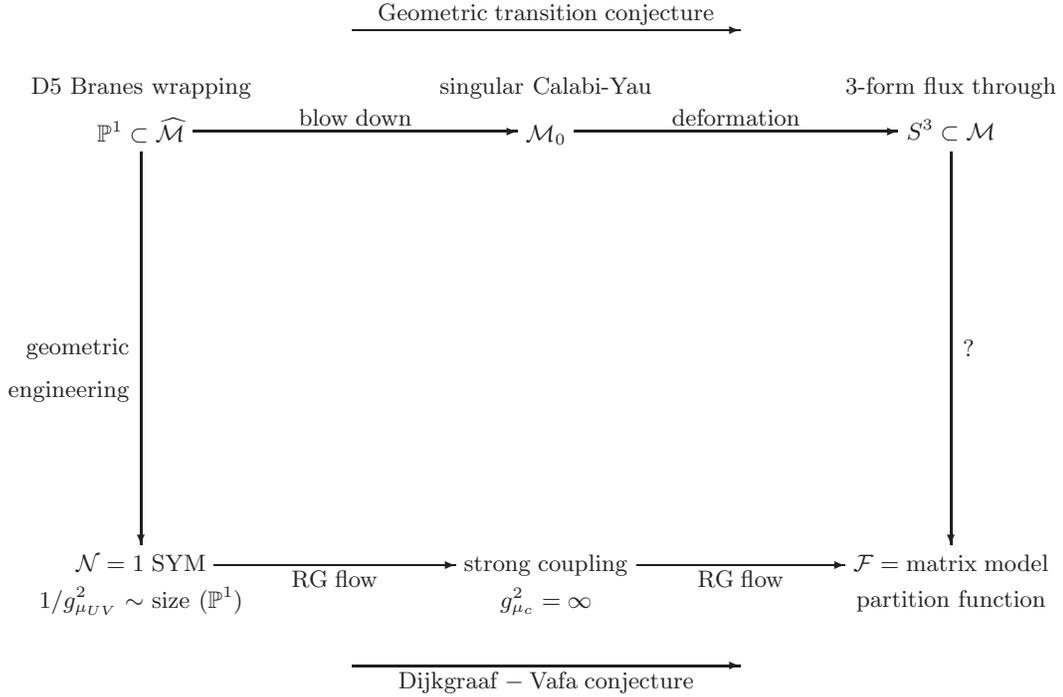

\begin{footnotesize}
$$\bfig
\putsquare<1`1`0`1;1400`1500>(0,250)[\PP^1 \subset \Hat{\M}`\M_0`\mathcal{N}=1 \mathrm{\;SYM}`\mathrm{strong\;coupling};
\mathrm{blow\;down}`\mathrm{geometric}``\mathrm{RG\;flow}]
\putsquare<1`0`1`1;1400`1500>(1400,250)[\phantom{\M_0}`S^3\subset\M`\phantom{\mathrm{strong\;coupling}}`\mathcal{F}=\mathrm{matrix\;model};\mathrm{deformation}``?`\mathrm{RG\;flow}]
\putmorphism(700,2100)(1,0)[``\mathrm{Geometric\;transition\;conjecture}]{1400}1a
\putmorphism(700,-100)(1,0)[``\mathrm{Dijkgraaf-Vafa\;conjecture}]{1400}1b
\put(0,1900){\makebox(0,0){D5 Branes wrapping}}
\put(1400,1900){\makebox(0,0){singular Calabi-Yau}}
\put(2800,1900){\makebox(0,0){3-form flux through}}
\put(0,125){\makebox(0,0){$1/g_{\mu_{UV}}^2 \sim$ size $(\PP^1)$}}
\put(1400,125){\makebox(0,0){$g_{\mu_c}^2 = \infty$}} 
\put(2800,125){\makebox(0,0){partition function}}
\put(-250,850){\makebox(0,0){engineering}}
\efig
$$
\end{footnotesize}
\caption{The big picture}
\label{fig:bigpicture}
\end{figure}

Recently, there has been a tremendous amount of work surrounding dualities which relate string theories to other classes of theories.  Maldacena's 1997 AdS/CFT correspondence is perhaps the most famous example of such a duality \cite{maldacena}.  The connection between Chern-Simons gauge theory and string theory was first introduced by Witten in 1992 \cite{witten}. In 1999, Gopakumar and Vafa initiated a program to study the relationship between large N limits of Chern-Simons theory (gauge theory) and type IIA topological string theory (geometry) by using ideas originally proposed by 't Hooft in the 1970's \cite{gopakumar}.  The resulting gauge theory/geometry correspondence led to a conjecture about extremal transitions, often referred to as the ``geometric transition conjecture.''  In the case of conifold singularities, this is more or less understood.  The conifold singularity can be resolved in two very different ways:  (1) with a traditional blow up in algebraic geometry, in which the singular point gets replaced by an exceptional $\PP^1$, or (2) by a deformation of the algebraic equation which replaces the singular point with an $S^3$ whose size is controlled by the deformation parameter (see Figure~\ref{fig:bigpicture}).  The physical degrees of freedom associated to D5 branes wrapping the $\PP^1$ correspond to a 3-form flux through the $S^3$.  The geometric statement is that one can interpolate between the two kinds of resolutions.

In 2002, Dijkgraaf and Vafa expanded this program and proposed new dualities between type IIB topological strings on Calabi-Yau threefolds and matrix models \cite{vafa1,vafa2}.  Due to the symmetry between type IIA and type IIB string theories, this may be viewed as ``mirror'' to the Gopakumar-Vafa conjecture.  By studying the conifold case, they found strong evidence for the matching of the string theory partition function with that of a matrix model whose potential is closely related to the geometry in question.  In particular, a dual version of special geometry in Calabi-Yau threefolds is seen in the eigenvalue dynamics of the associated matrix model \cite{vafa1}. The proposed string theory/matrix model duality has led to an explosion of research on matrix models, a topic which had been dormant since the early 1990's, when it was studied in the context of 2D gravity \cite{ginsparg}.  The connection between string theory and matrix models is of very tangible practical importance, since many quantities which are difficult to compute in string theory are much easier to handle on the matrix model side.

Inspired by these developments (summarized in Figure~\ref{fig:bigpicture}), in 2003 F. Ferrari was led to propose a direct connection between matrix models and the Calabi-Yau spaces of their dual string theories \cite{ferrari}.  It is well known that the solution to a 1-matrix model can be characterized geometrically, in terms of a hyperelliptic curve.  The potential for the matrix model serves as direct input into the algebraic equation for the curve, and the vacuum solutions (distributions of eigenvalues) can be obtained from the geometry of the curve and correspond to branch cuts on the Riemann surface.  The work of Vafa and collaborators on the strongly coupled dynamics of four-dimensional $\mathcal{N}=1$ supersymmetric gauge theories \cite{gopakumar,cachazo2,cachazo,vafa3} suggests that for multi-matrix models, higher-dimensional Calabi-Yau spaces might be useful. Ferrari pursues this idea in \cite{ferrari}, finding evidence that certain multi-matrix models can, indeed, be directly characterized in terms of higher-dimensional (non-compact) Calabi-Yau spaces.

By thinking of the matrix model potential $W(x_1,...,x_M)$ as providing constraints on the deformation space of an exceptional $\PP^1$ within a smooth (resolved) Calabi-Yau $\Hat{\M}$, Ferrari outlines a precise prescription for constructing such smooth geometries directly from the potential.  Specifically, the resolved geometry $\Hat{\M}$ is given by transition functions
\begin{equation}\label{eq:transfns} 
\beta = 1/\gamma, \skop v_1=\gamma^{-n}w_1, \skop
v_2=\gamma^{-m}w_2+\partial_{w_1}E(\gamma,w_1),
\end{equation}
between two coordinate patches $(\gamma,w_1,w_2)$ and $(\beta,v_1,v_2)$, where $\beta$ and $\gamma$ are
stereographic coordinates over an exceptional $\PP^1$. 
The perturbation comes from the ``geometric potential'' $E(\gamma,w)$, which is related to the matrix
model potential $W$ via
\begin{equation}\label{eq:superpotential}
W(x_1,...,x_M)=\dfrac{1}{2\pi i}\oint_{C_0}
\gamma^{-M-1}E(\gamma,\sum_{i=1}^{M}x_i\gamma^{i-1})\mathrm{d}\gamma,
\end{equation}
where $M = n+1 = -m-1.$ We explain this construction in detail in Section~\ref{sec:ferrari}.

In the absence of the perturbation term $\partial_{w_1}E(\gamma,w_1)$, the transition functions~\eqref{eq:transfns} simply describe an $\OO(M-1) \oplus \OO(-M-1)$ bundle over a $\PP^1$.  The matrix model superpotential $W$ encodes the constraints on the sections $x_1,...,x_M$ of the bundle due to the presence of $\partial_{w_1}E(\gamma,w_1)$.
Note that this procedure is also invertible.  In other words, given a matrix model superpotential
$W(x_1,...,x_M)$, one can find a corresponding geometric potential $E(\gamma,w_1)$.  However, not all perturbation terms $\gamma^j w_1^k$ contribute to the superpotential~\eqref{eq:superpotential}, so there may be many choices of geometric potential for a given $W$. 
Nevertheless, the associated geometry $\Hat{\M}$ is unique \cite[page 634]{ferrari}.  

In 2000, S. Katz had already shown how to codify constraints on versal deformation spaces of curves in terms of a potential function,\footnote{For a rigorous derivation of the D-brane superpotential, see \cite{aspinwall}.} in cases such as~\eqref{eq:transfns} where the constraints are integrable \cite{katz}.  In 2001, F. Cachazo, S. Katz and C. Vafa constructed $\mathcal{N}=1$ supersymmetric gauge theories corresponding to D5 branes wrapping 2-cycles of ADE fibered threefolds \cite{cachazo}. Ferrari studies the same kinds of geometries in a different context, by interpreting the associated potential as belonging to a matrix model, and proposing that the Calabi-Yau geometry encodes all relevant information about the matrix theory.\footnote{Specifically, it is the triple of spaces $\Hat{\M},\M_0,$ and $\M$ that are conjectured to encode the matrix model quantities; the blow-down map $\pi:\Hat{\M} \longrightarrow \M_0$ is the most difficult step towards performing the matrix model computations \cite{ferrari}.}  He is able to verify this in a few examples, and computes known resolvents of matrix models in terms of periods in the associated geometries.  The matching results, as well as Ferrari's solution of a previously unsolved matrix model, suggest that not only can matrix models simplify computations in string theory, but associated geometries from string theory can simplify computations in matrix models.

Many questions immediately arise from Ferrari's construction.  In particular, the matrix model resolvents are not directly encoded in the resolved geometry $\Hat{\M}$, but require knowing the corresponding singular geometry $\M_0$ obtained by blowing down the exceptional $\PP^1$.  It is not clear how to do this blow-down in general. It is also not obvious that a geometry constructed from a matrix model potential in this manner will indeed contain a $\PP^1$ which can be blown down to become an isolated singularity.\footnote{We will see later that the `hat' potentials from the Intriligator-Wecht classification lead to geometries where an entire family of $\PP^1$'s is blown down to reveal a space $\M_0$ with non-isolated singularities.  It is interesting to wonder what the corresponding ``geometric transition conjecture'' should be for these cases.}  Just which matrix models can be ``geometrically engineered'' in this fashion?  What are the corresponding geometries?  Can different matrix model potentials correspond to the same geometry?  If so, what common features of those models does the geometry encode?  Ferrari asks many such questions at the end of his paper \cite{ferrari}, and also wonders whether or not it might be possible to devise an algorithm which will automatically construct the blow-down given the initial resolved space.

Previously established results in algebraic geometry such as Laufer's Theorem \cite{laufer} and the classification of Gorenstein threefold singularities by S. Katz and D. Morrison \cite{morrison} provide a partial answer to these questions.

\begin{theorem}[Laufer 1979]\label{thm:laufer}
Let $\M_0$ be an analytic space of dimension $D \geq 3$ with an isolated singularity at $p$.  Suppose there exists a non-zero holomorphic $D$-form $\Omega$ on $\M_0-\{p\}$.\footnote{This is the Calabi-Yau condition.} Let $\pi:\Hat{\M} \longrightarrow \M_0$ be a resolution of $\M_0$.  Suppose that the exceptional set $A=\pi^{-1}(p)$ is one-dimensional and irreducible. Then $A \cong \PP^1$ and $D=3$.
Moreover, the normal bundle of $\PP^1$ in $\Hat{\M}$ must be either $\mathcal{N} = \OO(-1)\oplus\OO(-1),
\OO\oplus\OO(-2),$ or $\OO(1)\oplus\OO(-3)$. 
\end{theorem}

\noindent Laufer's theorem immediately tells us that we can restrict our search of possible geometries to dimension 3, and that there are only three candidates for the normal bundle to our exceptional $\PP^1$.  In Ferrari's construction, the bundles
$\OO(-1)\oplus\OO(-1), \OO\oplus\OO(-2),$ and $\OO(1)\oplus\OO(-3)$ correspond to zero-, one-, and two-matrix models, respectively.\footnote{This is because these bundles have zero, one, and two independent global sections, respectively.} Following the Dijkgraaf-Vafa correspondence, this puts a limit of 2 adjoint fields on the associated gauge theory, which is precisely the requirement for asymptotic freedom in $\N=1$ supersymmetric gauge theories.  This happy coincidence is perhaps our first indication that the Calabi-Yau geometry encodes information about the RG flow of its corresponding matrix model or gauge theory. 

The condition $M \leq 2$ for the normal bundle $\OO(M-1)\oplus\OO(-M-1)$ 
in Laufer's theorem is equivalent to asymptotic freedom, and reflects the fact that only for asymptotically
free theories can we expect the $\PP^1$ to be exceptional.  In considering matrix model potentials with 
$M \geq 3$ fields, Ferrari points out that the normal bundle to the $\PP^1$ changes with the addition
of the perturbation $\partial_{w_1}E(\gamma,w_1)$, and makes the following conjecture \cite[page 636]{ferrari}:

\begin{conjecture}[RG Flow, Ferrari 2003]  
Consider the perturbed geometry for $m=-n-2$ and associated superpotential $W$.  Let $\N$ be the normal bundle of a $\PP^1$ that sits at a given critical point of $W$.
Let $r$ be the corank of the Hessian of $W$ at the critical point.  Then
$\N = \OO(r-1)\oplus\OO(-r-1)$.
\end{conjecture}

\noindent Ferrari proves the conjecture for $n=1$, and limits himself to two-matrix models
($M=n+1=2$) in the rest of his paper.  Our first result gives evidence in support of the RG flow conjecture
in a more general setting.\footnote{It has since come to our attention that the full conjecture
has been proven in \cite{ricco}.}

\begin{myproposition} For $-M \leq r \leq M$, the addition of the perturbation term $\partial_{w_1}E(\gamma,w_1)=\gamma^{r+1}w_1$ in the transition functions
$$\beta = \gamma^{-1}, \skop v_1 = \gamma^{-M+1}w_1, \skop v_2 =
\gamma^{M+1}w_2+\gamma^{r+1}w_1,$$ changes the bundle from
$\OO(M-1)\oplus\OO(-M-1)$ to $\OO(r-1)\oplus\OO(-r-1).$
In particular, the $M$--matrix model potential
$$W(x_1,...,x_M) = \dfrac{1}{2}\sum_{i=1}^{M-r}x_ix_{M-r+1-i}, \skop (r \geq 0)$$
is geometrically equivalent \footnote{We will call two potentials {\em geometrically equivalent} if they yield the same geometry under Ferrari's construction.} 
to the $r$--matrix model potential
$$W(x_1,...,x_r) = 0.$$
\end{myproposition}

\noindent The proof is given in Section~\ref{sec:bundlechange}.  The fact that the associated superpotential is purely quadratic is satisfying since for quadratic potentials we can often ``integrate out'' fields, giving a field--theoretic intuition for why the geometry associated to an $M$--matrix model can be equivalent to that of an $r$--matrix model, with $r < M$.

Laufer's theorem constrains the dimension, the exceptional set and its normal bundle--but what are the possible singularity types?  In the surface case (complex dimension two), the classification of simple singularities is a classic result.\footnote{An excellent reference for this and other results in singularity theory is \cite{arnold2}.  For a more applications-oriented treatment (with many cute pictures!) see Arnol$'$d's 1991 book \cite{arnold}.
For 15 characterizations of rational double points, see \cite{durfee}.}    As hypersurfaces in $\CC^3$, the distinct geometries are given by: 
\begin{equation}\label{dim2}
\begin{array}{ccccc}
A_k &: & x^2 + y^2 + z^{k+1} &=& 0\\
D_{k+2} &: & x^2 + y^2z + z^{k+1} &=& 0\\
E_6 &: & x^2 + y^3 + z^4 &=& 0\\
E_7 &: & x^2 + y^3 + yz^3 &=& 0\\
E_8 &: & x^2 + y^3 + z^5 &=& 0
\end{array}
\end{equation}

In 1992, Katz and Morrison answered this question in dimension 3 when they characterized the full set of Gorenstein threefold singularities with irreducible small resolutions using invariant theory \cite{morrison}.  In order to do the
classification, Katz and Morrison find it useful to think of threefolds as deformations of surfaces, where the deformation parameter $t$ takes on the role of the extra dimension.  The equations for the singularities can thus
be written in so-called {\em preferred versal form}, as given in Table~\ref{table:versal}.
The coefficients $\alpha_i, \delta_i, \gamma_i,$ and $\varepsilon_i$ are given by invariant polynomials, and are implicity functions of the deformation parameter $t$.  We will also find this representation of the singular threefolds useful in identifying what kinds of singular geometries we get upon blowing down resolved geometries.

\begin{table}[h]
\begin{center}
$\begin{array}{|c|r@{\  + \ }l|} \hline
 & \multicolumn{2}{c|}{} \\
\mathrm{S} & \multicolumn{2}{c|}{\text{Preferred Versal Form}}  \\
 & \multicolumn{2}{c|}{} \\ \hline
 & \multicolumn{2}{c|}{} \\
A_{n-1}   &  - X Y + Z^n  &   \sum_{i=2}^{n}\a_i Z^{n-i} \\
(n \ge 2) & \multicolumn{2}{c|}{} \\
  & \multicolumn{2}{c|}{} \\
D_n  &  \multicolumn{1}{r@{\  - \ }}{ X^2 + Y^2 Z - Z^{n-1} } &
 \sum_{i=1}^{n-1}\d_{2i} Z^{n-i-1}  + 2 \c_n Y  \\
(n \ge 3)  & \multicolumn{2}{c|}{} \\
  & \multicolumn{2}{c|}{} \\
E_4   &  - X Y + Z^5  &   \e_2 Z^3 + \e_3 Z^2 + \e_4 Z + \e_5 \\
  & \multicolumn{2}{c|}{} \\
E_5  &  \multicolumn{1}{r@{\  - \ }}{  X^2 + Y^2 Z - Z^4 } &
 \e_2 Z^3 - \e_4 Z^2  + 2 \e_5 Y - \e_6 Z - \e_8  \\
  & \multicolumn{2}{c|}{} \\
E_6  &  - X^2 - X Z^2 + Y^3  &
      \e_2 Y Z^2 + \e_5 Y Z + \e_6 Z^2 + \e_8 Y \\
  &  &
      \e_9 Z + \e_{12} \\
  & \multicolumn{2}{c|}{} \\
E_7  &  - X^2 - Y^3 + 16 Y Z^3  &
    \e_2 Y^2 Z + \e_6 Y^2 + \e_8 Y Z + \e_{10} Z^2  \\
  &  &
       \e_{12} Y
     + \e_{14} Z + \e_{18} \\
  & \multicolumn{2}{c|}{} \\
E_8  &  - X^2 + Y^3 - Z^5 &
      \e_2 Y Z^3 + \e_8 Y Z^2 + \e_{12} Z^3 + \e_{14} Y Z \\
  &  &
      \e_{18} Z^2 + \e_{20} Y + \e_{24} Z
     + \e_{30} \\
 & \multicolumn{2}{c|}{} \\ \hline
\end{array}$
\end{center}
\caption[Gorenstein threefold singularities in preferred versal form]{Gorenstein threefold singularities in preferred versal form \protect\cite[page 465]{morrison}.}
\label{table:versal}
\end{table}

In contrast to what one might expect,\footnote{By taking hyperplane sections, one may get surface singularities corresponding to any of the ADE Dynkin diagrams.  A priori, this could indicate that there is an infinite number of families of the threefold singularities with irreducible small resolutions.  What Katz and Morrison discovered is that only a finite number of Dynkin diagrams can arise from ``generic'' hyperplane sections.} there are only a finite number of families of Gorenstein threefold singularities with irreducible small resolutions.  They are distinguished by the Koll\'ar ``length'' invariant,\footnote{\textbf{Defn:} Let $\pi: Y \flecha X$ irreducible small resolution of an isolated threefold singularity $p \in X$.  Let $C = \pi^{-1}(p)$ be the exceptional set.  The {\em length} of $p$ is the length at the generic point of the scheme supported on $C$, with structure sheaf $\OO_Y/\pi^{-1}(m_p,x)$.} and are resolved via small resolution of the appropriate length node in the corresponding Dynkin diagram.  The precise statement of Katz and Morrison's results are given by the following theorem and corollary \cite[page 456]{morrison}:

\begin{theorem}[Katz \& Morrison 1992] The generic hyperplane section of an isolated Gorenstein threefold singularity which has an irreducible small resolution defines one of the primitive partial resolution graphs in Figure~\ref{figure1}.  Conversely, given any such primitive partial resolution graph, there exists an
irreducible small resolution $Y\to X$ whose general hyperplane
section is described by that partial resolution graph.
\end{theorem}

\begin{corollary}  The general hyperplane section of $X$ is uniquely determined
by the length of the singular point $P$.
\end{corollary}

\setlength{\unitlength}{.75 in}

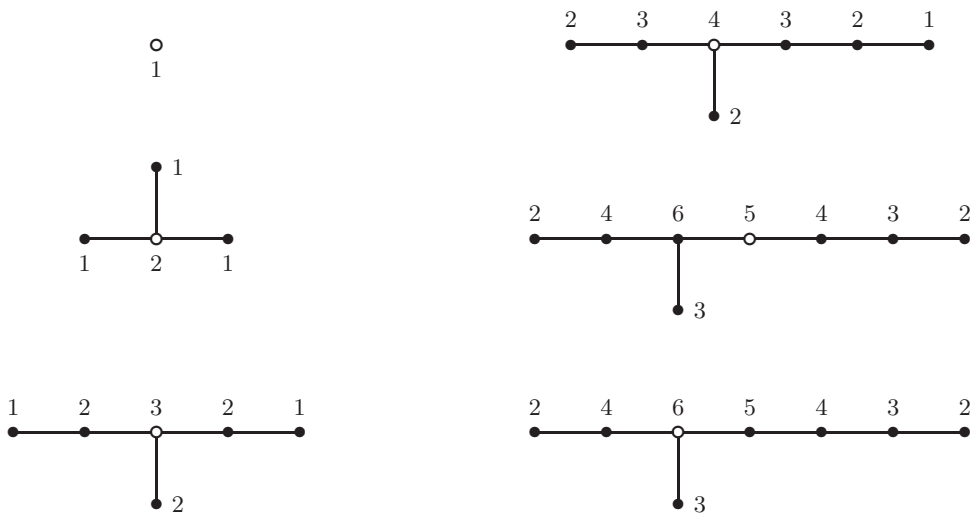
\begin{figure}[t]
\begin{picture}(2,1)(1.9,.5)
\thicklines
\put(2.9,1){\circle{.075}}
\put(2.775,.7){\makebox(.25,.25){\footnotesize 1}}
\end{picture}
\hspace*{\fill}
\begin{picture}(3,1)(1.65,.5)
\thicklines
\put(1.9,1){\circle*{.075}}
\put(1.9,1){\line(1,0){.5}}
\put(2.4,1){\circle*{.075}}
\put(2.4,1){\line(1,0){.4625}}
\put(2.9,1){\circle{.075}}
\put(2.9,.9625){\line(0,-1){.4625}}
\put(2.9,.5){\circle*{.075}}
\put(2.9375,1){\line(1,0){.4625}}
\put(3.4,1){\circle*{.075}}
\put(3.4,1){\line(1,0){.5}}
\put(3.9,1){\circle*{.075}}
\put(3.9,1){\line(1,0){.5}}
\put(4.4,1){\circle*{.075}}
\put(1.775,1.05){\makebox(.25,.25){\footnotesize 2}}
\put(2.275,1.05){\makebox(.25,.25){\footnotesize 3}}
\put(2.775,1.05){\makebox(.25,.25){\footnotesize 4}}
\put(2.925,.375){\makebox(.25,.25){\footnotesize 2}}
\put(3.275,1.05){\makebox(.25,.25){\footnotesize 3}}
\put(3.775,1.05){\makebox(.25,.25){\footnotesize 2}}
\put(4.275,1.05){\makebox(.25,.25){\footnotesize 1}}
\end{picture}

\bigskip

\bigskip

\begin{picture}(2,1)(1.9,.5)
\thicklines
\put(2.4,1){\circle*{.075}}
\put(2.4,1){\line(1,0){.4625}}
\put(2.9,1){\circle{.075}}
\put(2.9,1.0375){\line(0,1){.4625}}
\put(2.9,1.5){\circle*{.075}}
\put(2.9375,1){\line(1,0){.4625}}
\put(3.4,1){\circle*{.075}}
\put(2.275,.7){\makebox(.25,.25){\footnotesize 1}}
\put(2.775,.7){\makebox(.25,.25){\footnotesize 2}}
\put(2.925,1.375){\makebox(.25,.25){\footnotesize 1}}
\put(3.275,.7){\makebox(.25,.25){\footnotesize 1}}
\end{picture}
\hspace*{\fill}
\begin{picture}(3,1)(1.9,.5)
\thicklines
\put(1.9,1){\circle*{.075}}
\put(1.9,1){\line(1,0){.5}}
\put(2.4,1){\circle*{.075}}
\put(2.4,1){\line(1,0){.5}}
\put(2.9,1){\circle*{.075}}
\put(2.9,1){\line(0,-1){.5}}
\put(2.9,.5){\circle*{.075}}
\put(2.9,1){\line(1,0){.4625}}
\put(3.4,1){\circle{.075}}
\put(3.4375,1){\line(1,0){.4625}}
\put(3.9,1){\circle*{.075}}
\put(3.9,1){\line(1,0){.5}}
\put(4.4,1){\circle*{.075}}
\put(4.4,1){\line(1,0){.5}}
\put(4.9,1){\circle*{.075}}
\put(1.775,1.05){\makebox(.25,.25){\footnotesize 2}}
\put(2.275,1.05){\makebox(.25,.25){\footnotesize 4}}
\put(2.775,1.05){\makebox(.25,.25){\footnotesize 6}}
\put(2.925,.375){\makebox(.25,.25){\footnotesize 3}}
\put(3.275,1.05){\makebox(.25,.25){\footnotesize 5}}
\put(3.775,1.05){\makebox(.25,.25){\footnotesize 4}}
\put(4.275,1.05){\makebox(.25,.25){\footnotesize 3}}
\put(4.775,1.05){\makebox(.25,.25){\footnotesize 2}}
\end{picture}

\bigskip

\bigskip

\begin{picture}(2,1)(1.9,.5)
\thicklines
\put(1.9,1){\circle*{.075}}
\put(1.9,1){\line(1,0){.5}}
\put(2.4,1){\circle*{.075}}
\put(2.4,1){\line(1,0){.4625}}
\put(2.9,1){\circle{.075}}
\put(2.9,.9625){\line(0,-1){.4625}}
\put(2.9,.5){\circle*{.075}}
\put(2.9375,1){\line(1,0){.4625}}
\put(3.4,1){\circle*{.075}}
\put(3.4,1){\line(1,0){.5}}
\put(3.9,1){\circle*{.075}}
\put(1.775,1.05){\makebox(.25,.25){\footnotesize 1}}
\put(2.275,1.05){\makebox(.25,.25){\footnotesize 2}}
\put(2.775,1.05){\makebox(.25,.25){\footnotesize 3}}
\put(2.925,.375){\makebox(.25,.25){\footnotesize 2}}
\put(3.275,1.05){\makebox(.25,.25){\footnotesize 2}}
\put(3.775,1.05){\makebox(.25,.25){\footnotesize 1}}
\end{picture}
\hspace*{\fill}
\begin{picture}(3,1)(1.9,.5)
\thicklines
\put(1.9,1){\circle*{.075}}
\put(1.9,1){\line(1,0){.5}}
\put(2.4,1){\circle*{.075}}
\put(2.4,1){\line(1,0){.4625}}
\put(2.9,1){\circle{.075}}
\put(2.9,.9625){\line(0,-1){.4625}}
\put(2.9,.5){\circle*{.075}}
\put(2.9375,1){\line(1,0){.4625}}
\put(3.4,1){\circle*{.075}}
\put(3.4,1){\line(1,0){.5}}
\put(3.9,1){\circle*{.075}}
\put(3.9,1){\line(1,0){.5}}
\put(4.4,1){\circle*{.075}}
\put(4.4,1){\line(1,0){.5}}
\put(4.9,1){\circle*{.075}}
\put(1.775,1.05){\makebox(.25,.25){\footnotesize 2}}
\put(2.275,1.05){\makebox(.25,.25){\footnotesize 4}}
\put(2.775,1.05){\makebox(.25,.25){\footnotesize 6}}
\put(2.925,.375){\makebox(.25,.25){\footnotesize 3}}
\put(3.275,1.05){\makebox(.25,.25){\footnotesize 5}}
\put(3.775,1.05){\makebox(.25,.25){\footnotesize 4}}
\put(4.275,1.05){\makebox(.25,.25){\footnotesize 3}}
\put(4.775,1.05){\makebox(.25,.25){\footnotesize 2}}
\end{picture}

\caption{The 6 types of Gorenstein threefold singularities}
\label{figure1}
\end{figure}

We thus know that there are only a finite number of families of distinct geometries with the desired properties for Ferrari's construction, and they correspond to isolated threefold singularities with small resolutions. While much is known about the resolution of these singularities (they are obtained by blowing up divisors associated to nodes of the appropriate length in the corresponding Dynkin diagram), it is not easy to perform the small blowup explicitly.

The major obstacle in identifying which matrix model corresponds to each of the candidate singular geometries from \cite{morrison} is the absence of a simple description in the form of~\eqref{eq:transfns} for their small resolutions.  This frustration is also expressed in \cite{cachazo}, where the same geometries are used to construct $\mathcal{N}=1$ ADE quiver theories.\footnote{``...the gauge theory description suggests a rather simple global geometric description of the blown up $\PP^1$ for all cases.  However such a mathematical construction is not currently known in the full generality suggested by the gauge theory.  Instead only some explicit blown up geometries are known in detail...''\cite[page 35]{cachazo}}  In the case where the normal bundle to the exceptional $\PP^1$ is $\OO(1)\oplus\OO(-3)$, only Laufer's example \cite{laufer} and its extension by Pinkham and Morrison 
\cite[page 368]{pinkham} was known.  For us, the resolution to this problem came from a timely, albeit surprising, source.

In September, 2003, Intriligator and Wecht posted their results on RG fixed points of $\mathcal{N}=1$ SQCD with adjoints \cite{wecht}.  Using ``a-maximization'' and doing a purely field theoretic analysis, they classify all
relevant adjoint superpotential deformations for 4d $\N=1$ SQCD with
$\N_f$ fundamentals and $\N_a=2$ adjoint matter chiral superfields,
$X$ and $Y$.\footnote{For a related study of these theories, see \cite{mazzucato}.}  The possible RG fixed points, together with the map of possible flows between fixed points, are summarized in Figure 3.

\begin{figure}[h]
\label{fig:RGflow}
\begin{picture}(6,3)(0,0)
\put(.8,1.2){\makebox{
$\begin{array}{c|c}
\mathrm{type} & W(X,Y)\\
\hline
&\\
\Hat{O} & 0\\
\Hat{A} & \Tr Y^2\\
\Hat{D} & \Tr XY^2\\
\Hat{E} & \Tr Y^3\\
A_k & \Tr(X^{k+1}+Y^2)\\
D_{k+2} & \Tr(X^{k+1}+XY^2)\\
E_6 & \Tr(Y^3+X^4)\\
E_7 & \Tr(Y^3+YX^3)\\
E_8 & \Tr(Y^3+X^5)
\end{array}$}}
\put(3.7,0){\makebox{\includegraphics[scale=.7]{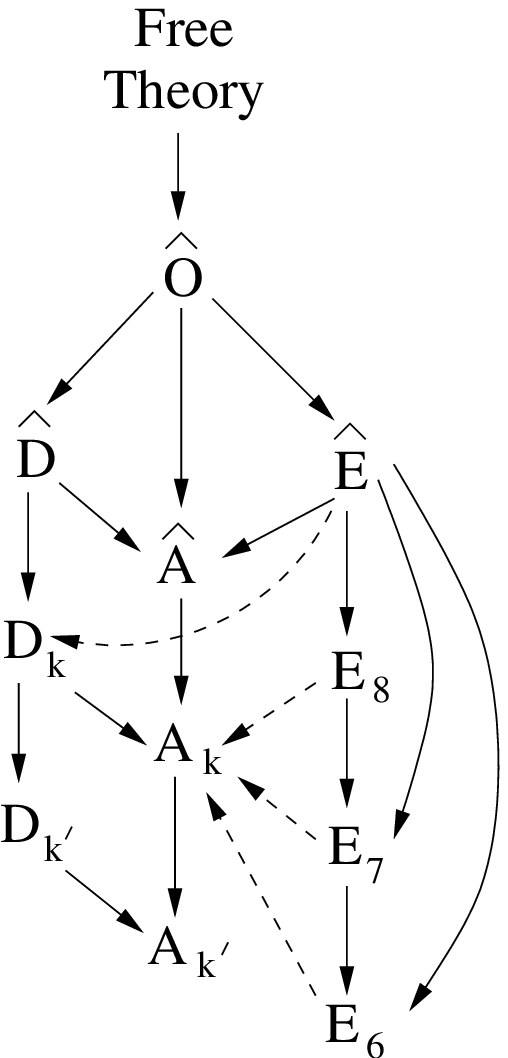}}}
\end{picture}
\caption[Intriligator--Wecht Classification of RG Fixed Points]{Intriligator--Wecht Classification 
of RG Fixed Points.  The diagram on the right shows the map of possible
flows between fixed points.  Dotted lines indicate flow to a particular value of $k$.  Note
that $k' < k$.\protect \cite[pages 3-4]{wecht}}
\end{figure}

Due to the form of the polynomials, Intriligator and Wecht name the relevant superpotential deformations according to the famous ADE classification of singularities in dimensions \nolinebreak 1 and 2 (see equation~\eqref{dim2}).  There is no geometry in their analysis, however, and they seem surprised to uncover a connection to these singularity types.\footnote{``On the face of it, this has no obvious connection to any of the other known ways in which Arnold's singularities have appeared in mathematics or physics.''\cite[page 3]{wecht}}

Naively, one may speculate that this is the answer.\footnote{In particular, if the Dijkgraaf--Vafa conjecture holds, we should expect any classification of $\N=1$ gauge theories to have a matrix model counterpart.  Verifying such a correspondence thus provides a non-trivial consistency check on the proposed string theory/matrix model duality.}  We make the following conjecture:

\begin{myconjecture}  The superpotentials in Intriligator and Wecht's ADE classification for $\N=1$ gauge theories (equivalently, the polynomials defining simple curve singularities), are precisely the matrix model potentials which yield small resolutions of Gorenstein threefold singularities using Ferrari's construction~\eqref{eq:transfns}.
\end{myconjecture}

\noindent Armed with this new conjecture, we may now run the classification program backwards.  Starting from the resolved space $\Hat{\M}$ given by transition functions over the exceptional $\PP^1$, we can verify the correspondence by simply performing the blow down and confirming that the resulting geometry has the right singularity type.  In particular, the matrix model superpotentials (if correct) give us simple descriptions for the small resolutions of Gorenstein threefold singularities in terms of transition functions as in~\eqref{eq:transfns}.  Like other geometric insights stemming from dualities in string theory, such a result is of independent mathematical interest.

This still leaves us with some major challenges.  As Ferrari pointed out, there was no known systematic way of performing the blow downs, and our first task was to devise an algorithm to do so \cite{thesis}.  The algorithm can be implemented by computer,\footnote{See \cite{thesis} for Maple code.} and searches for global holomorphic functions which can be used to construct the blow down.  Any global holomorphic function on the resolved geometry $\Hat{\M}$ is necessarily constant on the exceptional $\PP^1$, so these functions are natural candidates for coordinates on the blown down geometry $\M_0$, since the $\PP^1$ must collapse to a point.  The algorithm finds all (independent) global holomorphic functions which can be built from a specified list of monomials.  Because such a list can never be exhaustive, the resulting singular space $\M_0$, whose defining equations are obtained by finding relations among the global holomorphic functions, must be checked.  We can verify that we do, in fact, recover the original smooth space by inverting the blow down and performing the small resolution of the singular point.  Once we have shown that the collection of global holomorphic functions gives us the right blow-down map, it does not really matter how we found them.  Because it may be used more generally for finding blow-down maps (in particular, for resolved geometries corresponding to potentials we have not considered here), we include a description of the algorithm in the Appendix.

We find that this program works perfectly in the $A_k$ (length 1) and $D_{k+2}$ (length 2) cases, lending credence to the idea that the Intriligator--Wecht classification is, indeed, the right answer.  In the exceptional cases, however, a few mysteries arise.  We are only able to find the blow down for the Intriligator--Wecht superpotential $E_7$, and the resulting singular space has a length 3 singularity.  We summarize these results in the following theorem:

\begin{mytheorem}\label{thm1} 
Consider the 2-matrix model potentials $W(x,y)$ in Table~\ref{table:thm1}, with corresponding resolved geometries $\Hat{\M}$,
$$ \beta = 1/\gamma, \skop v_1=\gamma^{-1}w_1, \skop
v_2=\gamma^{3}w_2+\partial_{w_1}E(\gamma,w_1),$$
given by perturbation terms $\partial_{w_1}E(\gamma,w_1)$.  Blowing down the exceptional $\PP^1$ in each $\Hat{\M}$ yields the 
singular geometries $\M_0$ given in Table~\ref{table:thm1}.
\begin{table}[h]
$$\begin{array}{c|c|c|c}
\mathrm{type} & W(x,y) & \partial_{w_1}E(\gamma,w_1) & \mathrm{singular\;\;geometry\;\;} \M_0\\
\hline
&&&\\
A_k & \dfrac{1}{k+1}x^{k+1}+\12 y^2 & \gamma^2 w_1^k + w_1
& XY-T(Z^k-T) = 0\\
& & &\\
D_{k+2} & \dfrac{1}{k+1}x^{k+1}+xy^2 & \gamma^2 w_1^k+w_1^2 
& X^2 - Y^2Z + T(Z^{k/2} - T)^2 = 0, \:\:\;\; k \; \even\\
& & & X^2 - Y^2Z - T(Z^k - T^2) = 0, \:\:\;\; k \; \odd\\
& & &\\
E_7 & \dfrac{1}{3}y^3+yx^3 & \gamma^{-1}w_1^2+\gamma w_1^3
& X^2 - Y^3 + Z^5 + 3TYZ^2 + T^3Z = 0.\\
&&&\\
\hline
\end{array}$$
\caption{Superpotentials corresponding to length 1, length 2, and length 3 singularities.}
\label{table:thm1}
\end{table}
\end{mytheorem}

By comparing the above singular geometries with the equations in
preferred versal form (Table~\ref{table:versal}), we immediately identify the $A_k$ and $D_{k+2}$ superpotentials 
as corresponding to length 1 and length 2 threefold singularities, respectively.  For the $E_7$ potential, we first
note that the polynomial $X^2 - Y^3 + Z^5 + 3TYZ^2 + T^3Z$ is in preferred versal form for $E_8$ (with $\varepsilon_8 = -3T$ and $\varepsilon_{24}=-T^3$).\footnote{$T=0$ yields a hyperplane section with $E_8$ surface singularity.}  On the other hand, using Proposition 4 in the proof of the Katz--Morrison classification \cite[pages 499-500]{morrison}, we see that the presence of the monomial $T^3Z$ constrains the threefold singularity type to length 3.  We thus
have the following corollary:

\begin{corollary}
The resolved geometries $\Hat{\M}$ given by the 2-matrix model potentials $A_k$, $D_{k+2}$, and $E_7$ in Table~\ref{table:thm1} are small resolutions for length 1, length 2, and length 3 singularities, respectively.
\end{corollary}

Although simple descriptions of the form~\eqref{eq:transfns} were previously known for small resolutions of length 1 and length 2 Gorenstein threefold singularities (Laufer's example \cite{laufer} in the length 2 case), it is striking that in no other case such a concrete representation for the blowup was known.  Theorem 1, together with its Corollary, show a length 3 example where the small resolution also has an extremely simple form:
$$ \beta = 1/\gamma, \skop v_1=\gamma^{-1}w_1, \skop
v_2=\gamma^{3}w_2+\gamma^{-1}w_1^2+\gamma w_1^3.$$

The proof of Theorem~\ref{thm1} is given in Section~\ref{sec:thm1}. Missing are examples of length 4, length 5, and length 6 singularity types. For the moment, we are skeptical about whether or not these are describable using geometries that are simple enough to fit into Ferrari's framework.

In some sense the Intriligator--Wecht classification does not contain enough superpotentials; only length 1, 2, and 3 singularities appear to be included.  On the other hand, there are too many: the additional superpotentials  $\Hat{O},\Hat{A},\Hat{D}$ and $\Hat{E}$ have no candidate geometries corresponding to the Katz--Morrison classification of Gorenstein threefold singularities!  What kind of geometries do these new cases correspond to?  And what (if any) is their relation to the original ADE classification?  Using Ferrari's framework and our new algorithmic blow down methods we are able to identify the geometries corresponding to these extra `hat' cases.  We summarize the results in the following theorem:

\begin{mytheorem}\label{thm2} The singular geometries corresponding to the $\Hat{O},\Hat{A},\Hat{D}$ and $\Hat{E}$ cases in the Intriligator--Wecht classification of superpotentials are given in Table~\ref{table:thm2}.
\begin{table}[h]
$$\begin{array}{c|c|c|c}
\mathrm{type} & W(x,y) & \partial_{w_1}E(\gamma,w_1) & \mathrm{singular\;\;geometry\;\;} \M_0\\
\hline
&&&\\
\Hat{O} & 0 & 0 & \CC^3/\ZZ_3\\
& & &\\
\Hat{A} & \12 y^2 & \gamma^2 w_1 & \CC \times \CC^2/\ZZ_2\\
& & &\\
\Hat{D} & xy^2 & \gamma w_1^2 & X^2 + Y^2Z - T^3 = 0\\
& & &\\
\Hat{E} & \dfrac{1}{3} y^3 & \gamma^2 w_1^2 & \mathrm{Spec} (\CC[a,b,u,v]/\ZZ_2)/(b^4 - u^2 - av).\\
&&&\\
\hline
\end{array}$$
\caption{Geometries for the `Hat' cases.}\label{table:thm2}
\end{table}
\end{mytheorem}

The proof of Theorem~\ref{thm2} is the content of Section~\ref{sec:thm2}.  We find that the resolved geometries have full families of $\PP^1$'s which are blown down, and the resulting singular spaces have interesting relations to the ADE cases.  The $\Hat{A}$ geometry is a curve of $A_1$ singularities, while the equation for $\Hat{D}$ looks like the equations for $D_{k+2}$ where the $k$-dependent terms have been dropped.  The identification of new, related geometries obtained by combining Ferrari's framework with the Intriligator--Wecht classification turns out to be one of the most interesting parts of our story.  The presence of these extra geometries may have implications for the relevant string dualities; perhaps the geometric transition conjecture can be expanded beyond isolated singularities.

It is surprising that even in the $\Hat{O}$ case, with $W(x,y) = 0$ superpotential, the geometry is highly non-trivial.  In fact, we find that it is the $A_1, A_k,$ and $\Hat{A}$ cases which are, in some sense, the ``simplest.''  Although the descriptions for the resolved geometries $\Hat{\M}$ in these cases make it appear as though the normal bundles to the exceptional $\PP^1$'s are all $\OO(1)\oplus\OO(-3)$ (as required by a 2-matrix model potential $W(x,y)$), these geometries can all be described with fewer fields.  A straightforward application of Proposition 1 shows that the $\Hat{A}$ case is equivalent to a one-matrix model with $W(x)=0$ superpotential.  Similarly, we will see in Section~\ref{sec:length1}
that 
$$W(x,y) = \dfrac{1}{k+1}x^{k+1} + \dfrac{1}{2}y^2 \skop \and \skop W(x) = \dfrac{1}{k+1}x^{k+1}$$
are geometrically equivalent, so the $A_k$ (length 1) cases are also seen to correspond to 1-matrix models, where
the $y$ field has been ``integrated out.''  This is a relief because we know that the exceptional $\PP^1$ after blowing up an $A_k$ singularity should have normal bundle $\OO \oplus \OO(-2)$.  When $k=1$, Proposition 1 further reduces the geometry to that of a 0-matrix model (no superpotential possible!), showing that $A_1$ is the most trivial case, with normal bundle $\OO(-1)\oplus\OO(-1)$.\footnote{In contrast, the normal bundle in the $\Hat{O}$ case is truly $\OO(1)\oplus\OO(-3)$, showing that this geometry requires a 2-matrix model description.}  These results are summarized in Table~\ref{table:Acases}, and can be understood as evidence for Ferrari's RG Conjecture.
(Compare with Intriligator and Wecht's map of possible RG flows in Figure 3.)

\begin{table}[h]
$$\begin{array}{c|c|c|c}
\mathrm{type} & 2-\mathrm{matrix\;model} & 1-\mathrm{matrix\;model} & 0-\mathrm{matrix\;model}\\
\hline
& & &\\
\Hat{O} & W(x,y) = 0 & &\\
& & &\\
\Hat{A} & W(x,y) = \12 y^2 & W(x) = 0 & \\
& & &\\
A_k & W(x,y) = \dfrac{1}{k+1}x^{k+1}+\12 y^2 & W(x) = \dfrac{1}{k+1}x^{k+1} &\\
& & &\\
A_1 & W(x,y) = \dfrac{1}{2}x^{2}+\12 y^2 & W(x) = \dfrac{1}{2}x^{2} & W = 0\\
&&&\\
\hline
\end{array}$$
\caption{Geometrically equivalent superpotentials.}\label{table:Acases}
\end{table}

Our analysis also indicates that the names are well-chosen: the $\Hat{A}$ and $\Hat{D}$ geometries are closely related to their $A_{k}$ and $D_{k+2}$ counterparts, and the same might be true for the $\Hat{E}$ case.  The relationship between the $\Hat{A},\Hat{D}$ and $\Hat{E}$ geometries and the ADE singularities is worth exploring for purely geometric reasons.  To summarize, string dualities have told us to enlarge the class of geometries considered in \cite{morrison}, and have pointed us to closely related `limiting cases' of these geometries with non-isolated singularities.

\setlength{\unitlength}{1pt}

\section{The geometric framework}

\noindent Here we review Ferrari's construction of non-compact Calabi-Yau's from
matrix model superpotentials, following section 3 of \cite{ferrari}.
The main idea behind the geometric setup is that deformations of the exceptional $\PP^1$ in a resolved geometry $\Hat{\M}$ correspond to adjoint fields in the gauge theory \cite{cachazo}.  Alternatively, the deformation space for a $\PP^1$ wrapped by D-branes can be thought of in terms of matrix models.  The number $N$ of D-branes wrapping the $\PP^1$ gives the size of the matrices ($N \times N$), while the number $M$ of independent sections of the $\PP^1$ normal bundle gives the number of matrices (an $M$--matrix model).


Inspired by the string theory dualities, Ferrari develops a recipe to go straight from the matrix model to a Calabi-Yau space.  If the dualities hold, all of the matrix model quantities should be computable from the corresponding geometry.  In this way, Ferrari's prescription provides a non-trivial consistency check on the Gopakumar-Vafa and Dijkgraaf-Vafa conjectures.  Moreover, such a Calabi-Yau space provides a natural higher-dimensional analogue for the spectral (hyperelliptic) curve which encodes the solution to the hermitian one-matrix model.


For each theory, there are three relevant Calabi-Yau spaces:
the resolved Calabi-Yau $\Hat{\M}$, the singular Calabi-Yau $\M_0$, and 
the smooth deformed space $\M$.
In short, Ferrari's game consists of the following steps:
\begin{enumerate}
\item Start with an $M$-matrix model matrix model superpotential 
      $W(x_1,...,x_M)$ and construct a smooth Calabi-Yau $\Hat{\M}$.  
      The details of this construction are presented below.
\item Identify the exceptional $\PP^1$ in the resolved space $\Hat{\M}$.
\item Blow down the exceptional $\PP^1$ to get the singular $\M_0$.  The blow down map is 
      $\pi:\Hat{\M} \flecha \M_0$.
\item Perturb the algebraic equation for $\M_0$ to get the smooth deformed space $\M$.
\item From the triple of geometries, compute matrix model quantities (resolvents).
\item Use standard matrix model techniques (loop equations) to check answers in cases where
      the matrix model solution is known.
\end{enumerate}

\noindent In this framework, the matrix model superpotential is encoded in the transition functions defining
the resolved geometry $\Hat{\M}$.  Ferrari shows that a wide variety of matrix model superpotentials
arise in this fashion, and that matrix model resolvents can be computed directly from the geometry.
In other words, the solution to the matrix model is encoded in the corresponding triple of Calabi-Yau's.

The bottleneck to this program is Step 3, the construction of the blow-down map.  While Ferrari's
ad-hoc methods for constructing the blow-down are successful in his particular examples, he does not know
how to construct the blow down in general.  Moreover, it seems the calculation of the blow down map $\pi$ is essentially equivalent to solving the associated matrix model, and hence it would be very useful to have an
algorithm which computes $\pi$ \cite[page 655]{ferrari}.


In this paper, we are mainly concerned with Steps 1-3.  Our goal is to show how to compute the blow-down map in a large class of examples, and therefore to understand better which singular geometries $\M_0$ arise from matrix models in Ferrari's framework.  In the future, it would be nice to implement the deformation to $\M$ and also to compute matrix model quantities for our examples (Steps 4-6).  For the present, this is beyond our scope.\skp

\subsection{Step 1: Construction of resolved Calabi-Yau}
 
\noindent We now turn to Step 1, the construction of the ``upstairs'' resolved space $\Hat{\M}$ given the matrix
model superpotential $W(x_1,...,x_k)$.  $\Hat{\M}$ is given be transition functions between just two coordinate charts over an exceptional $\PP^1$: $(\beta,v_1,v_2)$ in the first chart, and $(\gamma,w_1,w_2)$ in the second chart.  $\beta$ and $\gamma$ should thought of as stereographic coordinates for the $\PP^1$, with $\beta = \gamma^{-1}$.  The other
coordinates $v_1,v_2$ and $w_1,w_2$ span the normal directions to the $\PP^1$, and have non-trivial transition functions.

We first discuss the case where $W = 0$, in which the Calabi-Yau is the total space of a vector bundle over
the exceptional $\PP^1$.  We then show how a simple deformation of the transition functions leads to constraints on
the sections of the bundle.  The independent sections $x_1,...,x_k$ correspond to matrix degrees of freedom ($k$ independent sections for a $k$-matrix model).  The constraints can be encoded in a potential $W(x_1,...,x_k)$.  When $W$ is non-zero, our geometry $\Hat{\M}$ is no longer a vector bundle -- if the total space were a vector bundle, the sections $x_1,...,x_k$ would be allowed to move freely and therefore satisfy no constraints.  We shall refer to geometries with $W \neq 0$ as ``deformed'' or ``constrained'' bundles.\skp

\noindent \textbf{Pure $\OO(n)\oplus\OO(m)$ bundle}\\
\noindent Consider the following $\Hat{\MM}$ geometry for $n \geq 0$ and $m < 0$:
$$\beta = \gamma^{-1}, \skop v_1=\gamma^{-n}w_1, \skop v_2=\gamma^{-m}w_2.$$
There is an $(n+1)$-dimensional family of $\PP^1$'s that sit at
$$w_1(\gamma)=\sum_{i=1}^{n+1} x_i \gamma^{i-1}, \skop w_2(\gamma)=0.$$
We have no freedom in the $w_2$ coordinate because $m < 0$ precludes $v_2(\beta)$
from being holomorphic whenever $w_2(\gamma)$ is.
$w_1(\gamma)$ and $w_2(\gamma)$ define globally holomorphic sections, and
in the $\beta$ coordinate patch become
$$v_1(\beta)=\sum_{i=1}^{n+1} x_i \beta^{n-i+1},\skop v_2(\beta)=0.$$
The parameters $x_i$ are precisely the fields in the associated superpotential,
and they span the versal deformation space of the $\PP^1$'s. In this case there are no
constraints on the $x_i$'s, which corresponds to the fact that the superpotential is
$$W(x_1,...,x_{n+1})=0.$$
The geometry $\Hat{\M}$ is the total space of a vector bundle, which we might refer to as a ``free'' bundle because
it is not constrained.\skp

\noindent \textbf{Deformed bundle; enter superpotential}\\
\noindent Now consider the deformed geometry (with $n \geq 0$ and $m < 0$):
$$\beta = \gamma^{-1}, \skop v_1=\gamma^{-n}w_1, \skop
v_2=\gamma^{-m}w_2+\partial_{w_1}E(\gamma,w_1),$$
where $E(\gamma,w_1)$ is a function of two complex variables which can be Laurent
expanded in terms of entire functions $E_i$,
$$E(\gamma,w_1) = \sum_{i=-\infty}^{\infty} E_i(w_1)\gamma^i.$$
We call $E$ the ``geometric potential.''
The most general holomorphic section $(w_1(\gamma),w_2(\gamma))$ of
the normal bundle $\N$ to the $\PP^1$'s still has
$$w_1(\gamma)=\sum_{i=1}^{n+1} x_i \gamma^{i-1}, \skop
v_1(\beta)=\sum_{i=1}^{n+1} x_i \beta^{n-i+1},$$
but in order to ensure that $v_2(\beta)$ is holomorphic, the $x_i$'s will
have to satisfy some constraints.  Since a holomorphic $w_2(\gamma)$ can
only cancel poles in $\beta^{-j}$ for $j \geq |m|$, the $x_i$'s must
satisfy $|m|-1$ constraints in order to cancel remaining lower-order poles introduced
by the perturbation.  Hence the versal deformation space of the $\PP^1$ is spanned
by $n+1$ parameters $x_i$ satisfying $|m|-1 = -m-1$ constraints.  

For the $\PP^1$ to be isolated we need $n+1 = -m-1$, and we denote this quantity
(the number of fields) by $M$.  The constraints are integrable, and
equivalent to the extremization $\d W=0$ of the corresponding superpotential
$W(x_1,...,x_M).$  The $\PP^1$'s then sit at the critical points of the
superpotential, in the sense that for critical values of the $x_i$'s,
the pair $(w_1(\gamma), w_2(\gamma))$ will be a global holomorphic section
defining a $\PP^1$.\skp

\noindent \textbf{Summary: General transition functions for $\Hat{\M}$}\\
The resolved geometry $\Hat{\M}$ is described by two coordinate patches $(\gamma,w_1,w_2)$ and
$(\beta,v_1,v_2)$, with transition functions
$$\beta = 1/\gamma, \skop v_1=\gamma^{-n}w_1, \skop
v_2=\gamma^{-m}w_2+\partial_{w_1}E(\gamma,w_1).$$
In the absence of the $\partial_{w_1}E(\gamma,w_1)$ term, this would simply be an $\OO(n) \oplus \OO(m)$
bundle over the $\PP^1$ parametrized by the stereographic coordinates $\gamma$ and $\beta$.
The perturbation comes from the ``geometric potential'' $E(\gamma,w)$, which can be expanded as
$$E(\gamma,w) = \sum_{i=-\infty}^{+\infty} E_i(w)\gamma^i.$$
\skp

\noindent \textbf{The superpotential}\\
The matrix model superpotential encodes the constraints on the sections $x_1,...,x_M$ due to
the presence of the perturbation term $\partial_{w_1}E(\gamma,w_1)$ in the defining transition functions
for $\Hat{\M}$.  It can be obtained directly from the geometric potential via
$$W(x_1,...,x_M)=\dfrac{1}{2\pi i}\oint_{C_0}
\gamma^{-M-1}E(\gamma,\sum_{i=1}^{M}x_i\gamma^{i-1})\d\gamma,$$
where $$M = n+1 = -m-1.$$
The contour integral is meant as a bookkeeping device; $C_0$ should be taken as
a small loop encircling the origin.  The integral is a compact notation used
by Ferrari to encode all of the constraints at once.  The general method for going from
transition function perturbation (geometric potential) to superpotential was first presented in \cite{katz}.

This procedure is also invertible.  In other words, given a matrix model superpotential
$W(x_1,...,x_M)$ one can find a corresponding geometric potential $E(\gamma,w_1)$, 
and therefore construct the associated geometry.  $E$ is not in general unique; from the expression
for $W$ one can see that terms can always be added to the geometric potential which will not contribute
to the residue of the integrand, and hence will not affect the superpotential.  Such terms have no
effect on the geometry, however.

Going from $W$ to $E$ is essentially the implementation of Step 1
in Ferrari's game.  We now turn to Step 2, which is to locate the exceptional $\PP^1$'s.\skp

\subsection{Step 2: Locating the $\PP^1$'s}

\noindent The first task in constructing the blow-down maps is figuring out 
where the $\PP^1$'s that we want to blow down are located.
We will mostly be interested in the $M=2$ case,
$$\beta=\gamma^{-1},\skop v_1=\gamma^{-1}w_1, \skop v_2=\gamma^{3}w_2+\partial_{w_1}E(\gamma,w_1),$$
where we always have
$$w_1(\gamma)=x+\gamma y, \skop v_1(\beta)=\beta x + y,$$
with $x$ and $y$ critical points of $W(x,y)$ at the $\PP^1$'s.  Depending
on the form of the perturbation $\partial_{w_1}E(\gamma,w_1)$, $w_2(\gamma)$
will be chosen to cancel poles of order $\geq 3$.  The requirement that
$v_2(\beta)$ be holomorphic will fix $x$ and $y$ values to be the same as
for the critical points of $W(x,y)$.\skp

\subsection{Step 3: Finding the blow-down map in Ferrari's examples}\label{step3}

\noindent As previously mentioned, Ferrari has no systematic way of constructing the blow-down map
$$\pi:\Hat{\M} \flecha \M_0.$$  He successfully finds the blow-down in several examples, however, through clever
but ad-hoc methods.  To see how Ferrari finds the blow-down, see the main examples from his paper \cite{ferrari}.\skp

\subsection{Example:  $A_k$}\label{sec:A_k}

We now illustrate Steps 1-3 in a simple example.
Consider the matrix model potential $$W(x)=\dfrac{1}{k+1}x^{k+1}.$$
Since there is only 1 field, 
the resolved geometry $\Hat{\M}$ is given by transition functions
$$\beta = \gamma^{-1}, \skop v_1 = w_1, \skop v_2 = \gamma^2 w_2 + \gamma w_1^k,$$
for an $\OO\oplus\OO(-2)$ bundle over the exceptional $\PP^1$.
To locate the $\PP^1$, we first note that 
$$w_1(\gamma)=x=v_1(\beta)$$ 
are the only holomorphic sections for the $\OO$ line bundle.  Substituting this into
the transition function for $v_2$ yields
$$v_2(\beta) = \beta^{-2} w_2(\gamma) + \beta^{-1}x^k,$$
which is only holomorphic if $x^k = w_2(\gamma) = 0$.  Therefore we have a single $\PP^1$ located at
$$w_1(\gamma)=w_2(\gamma)=0, \skop v_1(\beta)=v_2(\beta)=0.$$
Note that the position of the $\PP^1$ corresponds exactly to the critical point of the superpotential:
$$\d W = x^k dx = 0 \skop \implies \skop x^k = 0.$$

\noindent \textbf{The blow-down}\\  
\noindent To find the blow down map $\pi$, we must look for global holomorphic functions (which will necessarily be constant on the $\PP^1$).  We can immediately write down
\begin{eqnarray*}
\pi_1 &=& v_1 = w_1,\\
\pi_2 &=& v_2 = \gamma^2 w_2 + \gamma w_1^k,
\end{eqnarray*}
which are independent.  Moreover, notice the combination
$\beta v_2 - v_1^k = \gamma w_2$.  This gives us
\begin{eqnarray*}
\pi_3 &=& \beta v_2 - v_1^k = \gamma w_2,\\
\pi_4 &=& \beta^2 v_2 - \beta v_1^k = w_2.
\end{eqnarray*}
Since $\beta = \pi_4/\pi_3$, we have immediately from the definition of $\pi_3$ the relation
$$\M_0:\;\; \pi_3^2 = \pi_4\pi_2-\pi_3\pi_1^k.$$
This corresponds to an $A_k$ (length 1) singularity!\skp

\noindent \textbf{The blowup}\\ 
\noindent We check our computation by inverting the blow-down.  If we define
$$v_3 = \beta v_2 - v_1^k, \skop \mathrm{and} \skop w_3 = \gamma w_2 + w_1^k,$$ 
we can write
$$\begin{array}{ccccc}
\pi_1 &=& v_1 &=& w_1,\\
\pi_2 &=& v_2 &=& \gamma w_3,\\
\pi_3 &=& v_3 &=& \gamma w_2,\\
\pi_4 &=& \beta v_3 &=& w_2. 
\end{array}$$
In particular
$$\beta = \pi_4/\pi_3 = \gamma^{-1}.$$
This suggests that to recover the small resolution $\Hat{\M}$ we should blow up 
$$\pi_3 = \pi_4 = 0 \skop \mathrm{in} \skop \M_0:\;\; \pi_3^2 = \pi_4\pi_2-\pi_3\pi_1^k.$$
Denoting the $\PP^1$ coordinates by $[\beta:\gamma]$ and imposing
the relation $$\beta \pi_3 = \gamma \pi_4$$ in the blowup, we find in each chart 
$$\begin{array}{c|c}
(\gamma = 1)& (\beta = 1)\\
\hline
&\\
\pi_4 = \beta \pi_3 & \pi_3 = \gamma \pi_4\\
\pi_3 = \beta \pi_2 - \pi_1^k & \gamma^2\pi_4 = \pi_2 - \gamma\pi_1^k\\
&\\
(\beta,\pi_1,\pi_2) & (\gamma,\pi_1,\pi_4)
\end{array}$$

\noindent The transition functions between the two charts are easily found to be
$$\beta = \gamma^{-1}, \skop \pi_1 = \pi_1, \skop \pi_2 = \gamma^2 \pi_4 + \gamma \pi_1^k.$$
Identifying with the original coordinates, we find
$$\beta = \gamma^{-1}, \skop v_1 = w_1, \skop v_2 = \gamma^2 w_2 + \gamma w_1^k.$$
This is exactly what we started with!

Note that for $k=1$, we have a bundle-changing superpotential (see Section~\ref{sec:bundlechange}), and the normal bundle to the exceptional $\PP^1$ is $\OO(-1)\oplus \OO(-1)$ instead of $\OO \oplus \OO(-2)$.
\label{sec:ferrari}

\section{Superpotentials which change bundle structure}\label{sec:bundlechange}

In this section we describe a family of superpotentials which change the underlying bundle structure, thus proving Proposition 1:

\begin{myproposition2} For $-M \leq r \leq M$, the addition of the perturbation term $\partial_{w_1}E(\gamma,w_1)=\gamma^{r+1}w_1$ in the transition functions
$$\beta = \gamma^{-1}, \skop v_1 = \gamma^{-M+1}w_1, \skop v_2 =
\gamma^{M+1}w_2+\gamma^{r+1}w_1,$$ changes the bundle from
$\OO(M-1)\oplus\OO(-M-1)$ to $\OO(r-1)\oplus\OO(-r-1).$
In particular, the $M$--matrix model potential
$$W(x_1,...,x_M) = \dfrac{1}{2}\sum_{i=1}^{M-r}x_ix_{M-r+1-i}, \skop (r \geq 0)$$
is geometrically equivalent \footnote{We will call two potentials {\em geometrically equivalent} if they yield the same geometry under Ferrari's construction.} 
to the $r$--matrix model potential
$$W(x_1,...,x_r) = 0.$$
\end{myproposition2}

Consider an $\OO(n)\oplus\OO(m)$ normal bundle over a $\PP^1$ with geometric potential
$$E(\gamma,w_1) = \frac{1}{2}\gamma^k w_1^2, \skop k \in \ZZ.$$
The perturbed transition functions are
$$\beta = 1/\gamma, \skop v_1 = \gamma^{-n} w_1, \skop v_2 = \gamma^{-m}w_2+\gamma^k w_1.$$
If $(n,m,k)$ satisfies
$$ -n \leq k \leq -m,$$
we can perform the following change of coordinates:
$$\begin{array}{cccccc}
\til{w_1} &=& w_1+\gamma^{-m-k}w_2, &\til{v_1} &=& v_2,\\
\til{w_2} &=& w_2, & \til{v_2} &=& -v_1+\beta^{n+k}v_2.
\end{array}$$
Notice that
\begin{eqnarray*}
\til{v_1} &=&  \gamma^{-m}w_2+\gamma^k w_1 = \gamma^k \til{w_1},\\
\til{v_2} &=& -\gamma^{-n} w_1 + \gamma^{-n-k}(\gamma^{-m}w_2+\gamma^k w_1) = \gamma^{-n-m-k}\til{w_2},
\end{eqnarray*}
and so the new transition functions are
$$\beta = 1/\gamma, \skop \til{v_1} = \gamma^{k}\til{w_1}, \skop \til{v_2} = \gamma^{-n-m-k}\til{w_2}.$$
The geometric potential has changed our $\OO(n) \oplus \OO(m)$ bundle into
an $\OO(-k)\oplus\OO(n+m+k)$ bundle, with no superpotential.
The corresponding superpotential can be computed for $n+m = -2$, and depends on
the number of fields $M = n+1 = -m -1$:

\begin{eqnarray*}
W(x_1,...,x_M) &=& \frac{1}{2\pi i} \oint_{C_0} \gamma^{-M-1} E(\gamma, \sum_{i=1}^M x_i\gamma^{i-1})\d\gamma\\
&=& \frac{1}{2}\sum_{i=1}^M x_i x_{M-k-i+2}
\end{eqnarray*}
Note that all of these bundle-changing superpotentials are purely quadratic!
In the cases of interest, where $n+m=-2,$ the condition for the change of coordinates to be valid becomes
$$-n \leq k \leq n+2.$$
For allowed pairs $(n,k)$ we can thus get $$\OO(n)\oplus\OO(-n-2) \flecha \OO(-k)\oplus\OO(k-2)$$
by means of the perturbation.  Alternatively, we can think of these examples as ``true'' $\OO(-k)\oplus\OO(k-2)$
bundles which can be rewritten to ``look like'' $\OO(n)\oplus\OO(-n-2)$ plus a superpotential term.
\skp

\noindent {\bf RG conjecture}\\
\noindent In order to make contact with Ferrari's RG conjecture (see Introduction), 
we change notation a bit from the previous discussion:
$$M = n+1, \skop r = k-1.$$
The perturbation term in the following transition functions
$$\beta = \gamma^{-1}, \skop w_1' = \gamma^{-M+1}w_1, \skop w_2' =
\gamma^{M+1}w_2+\gamma^{r+1}w_1,$$ changes the bundle
$$\OO(M-1)\oplus\OO(-M-1) \flecha \OO(r-1)\oplus\OO(-r-1),
\skop \for \skop -M \leq r \leq M.$$ The change of coordinates:
$$\begin{array}{cccccc}
v_1 &=& w_1+\gamma^{M-r}w_2, &v_1' &=& w_2',\\
v_2 &=& w_2, &v_2' &=& -w_1'+\beta^{M+r}w_2',
\end{array}$$
yields new transition functions
$$\beta = \gamma^{-1}, \skop v_1' = \gamma^{r+1}v_1, \skop v_2' = \gamma^{1-r}v_2.$$

\noindent {\bf The superpotential}\\
\noindent The superpotential corresponding to the perturbation
$$\partial_{w_1}E(\gamma,w_1)=\gamma^{r+1}w_1$$
is given by
$$W_r(x_1,...,x_M) = \left\{\begin{array}{c}
\displaystyle{\frac{1}{2}\sum_{i=1}^{M-r}x_ix_{M-r+1-i}}, \skop
\for \skop r \geq 0,\\
\\
\displaystyle{\frac{1}{2}\sum_{i=1-r}^{M}x_ix_{M-r+1-i}}, \skop \for
\skop r \leq 0.\end{array}\right.$$
This completes the proof of Proposition 1.

Notice that the case $r=M$ is not interesting, as the bundle remains
unchanged and the superpotential vanishes.  Moreover,
the symmetry $r \mapsto -r$ in the bundle expression interchanges
two different superpotentials, but this amounts to a simple change
of coordinates.  To see this, first note that:
$$\begin{array}{cccccc}
&r \geq 0: &W_r(x_1,...,x_M) &=
&\displaystyle{\frac{1}{2}\sum_{i=1}^{M-|r|}x_ix_{M-|r|+1-i}}&\\
&r \leq 0: &W_r(x_1,...,x_M) &=
&\displaystyle{\frac{1}{2}\sum_{i=1+|r|}^{M}x_ix_{M+|r|+1-i}}& =
\displaystyle{\frac{1}{2}\sum_{i=1}^{M-|r|}x_{i+|r|}x_{M+1-i}}\\
\end{array}$$
The direction of the coordinate shift depends on the sign of $r$:
$$\begin{array}{ccccc}
&r \geq 0:& r \mapsto -r & \mathrm{is \;\; equivalent \;\; to} & x_i
\mapsto x_{i+|r|}\\
&r \leq 0:& r \mapsto -r & \mathrm{is \;\; equivalent \;\; to} & x_i
\mapsto x_{i-|r|}
\end{array}$$
i.e. $r \mapsto -r$ on the bundle side is equivalent to a simple
coordinate change for the corresponding superpotential.

We summarize the first few examples in the following table:
$$\begin{array}{c|c|c|c|c|c|c|c}
r & -3 & -2 & -1 & 0 & 1 & 2 & 3 \\
\hline 
&&&&&&&\\
M=1 & & & & \12 x_1^2 & & &\\
M=2 & & & \12 x_2^2 & x_1 x_2 & \12 x_1^2 & &\\
M=3 & & \12 x_3^2 & x_2x_3 & x_1 x_3 + \12 x_2^2 & x_1 x_2 & \12 x_1^2 & \\
M=4 & \12 x_4^2 & x_3x_4 & x_2x_4+\12 x_3^2 & x_1x_4+x_2x_3 &
x_1x_3+ \12 x_2^2 & x_1 x_2 & \12 x_1^2
\end{array}$$

\noindent{\bf The Hessian}\\
\noindent We compute the partial derivatives of our bundle-changing
superpotentials:
$$\begin{array}{cccccccccc}
&r \geq 0:& \dfrac{\partial W_r}{\partial x_j} &=& x_{M-r+1-j},&
\dfrac{\partial^2 W_r}{\partial x_k \partial x_j} &=&
\delta_{k,M-r+1-j}& \for & 1 \leq j \leq M-r.\\
&&&&&&&&&\\
&r \leq 0:& \dfrac{\partial W_r}{\partial x_j} &=& x_{M-r+1-j},&
\dfrac{\partial^2 W_r}{\partial x_k \partial x_j} &=&
\delta_{k,M-r+1-j}& \for & 1-r \leq j \leq M.\\
\end{array}$$
In each case, there is only one $k$ for every $j$ which yields a
non-zero second-partial.  This means the Hessian matrix has at most
one non-zero entry in each row and in each column.  The corank of
the Hessian is thus easy to compute, and is equal to the number of
rows (or columns) comprised entirely of zeroes.  In both the $r \geq
0$ and $r \leq 0$ cases, the corank of the Hessian is $r$ (see the
ranges for $j$ values). This is consistent with what we expect from
Ferrari's RG conjecture.

\section{Proof of Theorem 1}\label{sec:thm1}
In this section we prove Theorem 1 by (i) computing the resolved geometry $\Hat{\M}$ corresponding to the Intriligator-Wecht superpotentials, (ii) finding global holomorphic functions (ghf's) to define a blow down map $\pi:\Hat{\M}\rightarrow \M_0$, (2) determining the geometry of the blow down $\M_0$ by finding relations among the ghf's, and (3) blowing up the singular point in $\M_0$ in order to check that we do, indeed, recover the original resolved space.  The global holomorphic functions are found using the algorithm described in the Appendix (see \cite{thesis} for a detailed implementation).  Techniques for performing the blow ups can be found in \cite{flops,thesis}.


\subsection{The Case $A_k$}

\noindent In this section we prove the first part of {\bf Theorem 1}: \\
\noindent {\em The resolved geometry corresponding
to the Intriligator--Wecht superpotential 
$$W(x,y) = \dfrac{1}{k+1}x^{k+1} + \dfrac{1}{2}y^2,$$
corresponds to the singular geometry
$$ XY-T(Z^k-T) = 0.$$}
\noindent In the next section we will also discover that this potential is geometrically equivalent to 
$$W(x) = \dfrac{1}{k+1}x^{k+1}.$$

\noindent {\bf The resolved geometry $\Hat{\MM}$}\\
\noindent From the Intriligator--Wecht superpotential
$$W(x,y) = \dfrac{1}{k+1}x^{k+1} + \dfrac{1}{2}y^2,$$
we compute the resolved geometry $\Hat{\MM}$ in terms of transition functions
$$\beta = \gamma^{-1}, \skop v_1 = \gamma^{-1} w_1, \skop v_2 = \gamma^3 w_2 + \gamma^2 w_1^k + w_1.$$
To find the $\PP^1$s, we substitute $w_1(\gamma) = x + \gamma y$ into the $v_2$ transition function
$$v_2(\beta) = \beta^{-3} w_2 + \beta^{-2}(x + \beta^{-1}y)^k + x + \beta^{-1}y.$$
If we choose
$$w_2(\gamma) = \dfrac{x^k - (x+\gamma y)^k}{\gamma},$$
in the $\beta$ chart the section is
$$v_1(\beta) = \beta x + y, \skop v_2(\beta) = \beta^{-2}x^k + \beta^{-1}y + x.$$
This is only holomorphic if
$$x^k = y = 0,$$
and so we have a single $\PP^1$ located at
$$w_1(\gamma) = w_2(\gamma) = 0, \skop v_1(\beta) = v_2(\beta) = 0.$$
This is exactly what we expect from computing critical points of the superpotential
$$\d W = x^k dx + y dy = 0.$$
\skp

\noindent{\bf Global holomorphic functions}\\
\noindent The transition functions
$$\beta = \gamma^{-1}, \skop v_1 = \gamma^{-1} w_1, \skop v_2 = \gamma^3 w_2 + \gamma^2 w_1^k + w_1,$$
are quasi-homogeneous if we assign the weights
$$\begin{array}{c|c|c|c|c|c}
\beta & v_1 & v_2 & \gamma & w_1 & w_2 \\
k-1 & k+1 & 2 & 1-k & 2 & 3k-1
\end{array}.$$
The global holomorphic functions will thus necessarily be quasi-homogeneous in these weights.
We find the following global holomorphic functions:
$$\begin{array}{c|ccccc}
2 & y_1 &=& v_2 &=& \gamma^3 w_2 + \gamma^2 w_1^k + w_1\\
k+1 & y_2 &=& \beta v_2 - v_1 &=& \gamma^2 w_2 + \gamma w_1^k\\
2k & y_3 &=& \beta^2 v_2-\beta v_1 &=& \gamma w_2 + w_1^k\\
3k-1 & y_4 &=& v_2^{k-1} v_1 - \beta^3 v_2 + \beta^2 v_1 &=&
\end{array}$$
These are the first 4 ``distinct'' functions produced by our algorithm, 
in the sense that none is contained in the ring generated by the other 3.
\skp

\noindent {\bf The singular geometry $\MM_0$}\\
\noindent We conjecture that the ring of 
global holomorphic functions is generated by $y_1,y_2,y_3$ and $y_4$, 
subject to the degree $4k$ relation
$$\MM_0: \;\;\; y_2 y_4 + y_3^2 + y_2^2 y_1^{k-1} - y_3 y_1^k = 0.$$
The functions $y_i$ give us a blow-down map whose image $\MM_0$
has an isolated $A_k$ singularity. To see this, consider the 
change of variables
$$\til{y}_4 = y_4 + y_2 y_1^{k-1}= \beta v_2^k - \beta^3 v_2 + \beta^2 v_1.$$
Note that like $y_4$, $\til{y}_4$ is also quasi-homogeneous of degree $3k-1$. 
The functions $y_1,y_2, y_3$ and $\til{y}_4$ now satisfy the simpler relation
$$\MM_0: \;\;\; y_2 \til{y}_4 + y_3(y_3 - y_1^k) = 0.$$
\skp

\noindent {\bf The blowup}\\
\noindent We now verify that we have identified the right singular space $\MM_0$ by inverting
the blow-down.  In the $\beta$ and $\gamma$ charts we find
$$\begin{array}{ccc|ccc}
\beta &=& y_3/y_2 & \gamma &=& y_2/y_3\\
v_1 &=& \beta y_1-y_2 = (y_3y_1 - y_2^2)/y_2 & w_1 &=& y_1 - \gamma y_2 = (y_1 y_3-y_2^2)/y_3\\
v_2 &=& y_1 & w_2 &=& \beta(y_3-w_1^k)\\
&&& &=& -\til{y}_4 + \dfrac{y_1^k - (y_1-\gamma y_2)^k}{\gamma}.
\end{array}$$
This suggests that we should blow up 
$$y_2 = y_3 = 0,$$
for the small resolution of $\MM_0$.  We introduce $\PP^1$ coordinates $[\beta:\gamma]$ such that
$\beta y_2 = \gamma y_3.$  The blow up in each chart is then
$$\begin{array}{c|c}
(\gamma = 1) & (\beta = 1)\\
&\\
y_3 = \beta y_2 & y_2 = \gamma y_3\\
\til{y_4} = \beta(y_1^k - \beta y_2) & y_3 = y_1^k - \gamma \til{y_4}\\
&\\
\mathrm{coords:\;\;} (y_1, y_2, \beta) & \mathrm{coords:\;\;}(y_1, \til{y_4}, \gamma)
\end{array}$$
\skp

\noindent{\bf Transition functions}\\ 
\noindent The transition functions between the $\beta$ and $\gamma$ charts are
$$\beta = \gamma^{-1},\skop y_1 = y_1, \skop y_2 = \gamma(y_1^k - \gamma \til{y}_4)
= -\gamma^2 \til{y}_4 + \gamma y_1^k.$$
Note that for $k > 1$, this is an $\OO\oplus\OO(-2)$ bundle over the exceptional $\PP^1$, and corresponds
to a superpotential with a single field ($M=1$):
$$W(x) = \dfrac{1}{k+1}x^{k+1}.$$
(For $k = 1$ the bundle is actually $\OO(-1)\oplus\OO(-1)$ and $W=0$.)

In terms of the original coordinates, the transition functions become
$$\beta = \gamma^{-1},\skop v_2 = \gamma^3 w_2 + \gamma^2 w_1^k + w_1, \skop
\beta v_2 - v_1 = \gamma^2 w_2 + \gamma w_1^k.$$
Substituting the second transition function into the third reveals $v_1 = \gamma^{-1} w_1,$ and so
we recover our original transition functions
$$\beta = \gamma^{-1}, \skop v_1 = \gamma^{-1} w_1, \skop v_2 = \gamma^3 w_2 + \gamma^2 w_1^k + w_1,$$
which define an $\OO(1)\oplus\OO(-3)$ bundle deformed by the two field ($M=2$) superpotential
$$W(x,y) = \dfrac{1}{k+1} x^{k+1} + \dfrac{1}{2}y^2.$$

\subsection{A puzzle}

\noindent \textbf{The problem}\\
\noindent We saw the $A_k$ case in Section~\ref{sec:A_k}, with geometry $\Hat{\M}$ given by the superpotential 
$$W(x) = \dfrac{1}{k+1}x^{k+1},$$
and hence corresponding to a deformed $\OO\oplus\OO(-2)$ bundle over the $\PP^1$, with one field. 
However, Intriligator and Wecht identify 
$$W(x,y) = \dfrac{1}{k+1}x^{k+1} + \dfrac{1}{2}y^2,$$
as corresponding to an $A_k$-type singularity, with an extra field $y$ which requires
that the transition functions look like
$$\beta = \gamma^{-1},\skop v_1 = \gamma^{-1} w_1, \skop v_2 = \gamma^3 w_2 + \gamma^2 w_1^k + w_1.$$
In particular, the geometry $\Hat{\M}$ looks like that of an $\OO(1)\oplus\OO(-3)$ bundle over the
exceptional $\PP^1$!  What's going on here?\skp

\noindent \textbf{Resolution of the problem}\\
\noindent For $n=1$ and $k=0$, Proposition 1 tells us that the superpotential
$$W(x,y) = \dfrac{1}{2} y^2$$
changes the bundle 
$$\OO(1)\oplus\OO(-3) \flecha \OO\oplus\OO(-2).$$
Hence the extra field $y$ from the Intriligator--Wecht potential (with purely quadratic contribution to the superpotential) can be ``integrated out.''  Its effect is to change the bundle for $A_k$ from $\OO(1)\oplus\OO(-3)$,
which is necessary for a two-field description, to reveal the true underlying $\OO\oplus\OO(-2)$ structure.
In other words, 
$$W(x,y) = \dfrac{1}{k+1}x^{k+1} + \dfrac{1}{2}y^2 \skop \and \skop W(x) = \dfrac{1}{k+1}x^{k+1}$$
are geometrically equivalent.\skp

\noindent {\bf Beyond Proposition 1}\\
\noindent Note that this is not just a straightforward application of Proposition 1, which implies that $W(x,y) = \dfrac{1}{2}y^2$ and $W(x) = 0$ are geometrically equivalent.  Beginning with transition functions for the 
Intriligator--Wecht $A_k$ superpotential $W(x,y) = \dfrac{1}{k+1}x^{k+1} + \dfrac{1}{2}y^2,$
$$\beta = \gamma^{-1}, \skop v_1 = \gamma^{-1} w_1, \skop v_2 = \gamma^3 w_2 + \gamma^2 w_1^k + w_1,$$
the change of coordinates suggested in the proof of Proposition 1 
$$\begin{array}{cccccc}
\til{v_1} &=& v_2, & \til{w_1} &=& w_1 + \gamma^3 w_2, \\
\til{v_2} &=& -v_1 + \beta v_2, & \til{w_2} &=& w_2,
\end{array}$$
does not yield the appropriate new transition functions.  Instead, the more complicated change of coordinates
$$\begin{array}{cccccc}
\til{v_1} &=& v_2, & \til{w_1} &=& w_1 + \gamma^3 w_2 + \gamma^2 w_1^k, \\
\til{v_2} &=& -v_1 + \beta v_2, & \til{w_2} &=& w_2 - \gamma^{-1}\left[(\gamma^3w_2 + \gamma^2w_1^k + w_1)^k-w_1^k\right],
\end{array}$$
is needed to give new transition functions
$$\beta = \gamma^{-1}, \skop \til{v_1} = \til{w_1}, \skop \til{v_2} = \gamma^2\til{w_2} + \gamma \til{w_1}^k,$$
corresponding to the superpotential $W(x) = \dfrac{1}{k+1}x^{k+1}.$

It would be interesting to try to generalize Proposition 1 to include examples such as this, where there
are additional terms in the superpotential besides the quadratic pieces which suggest a change in bundle structure.  Trying to understand what makes the change of coordinates possible in this case may give hints as to how the geometric picture for RG flow might be extended.  The ultimate goal would be to understand how ``integrating out'' the $y$ coordinate in a potential of the form $W(x,y) = f(x,y) + y^2$ affects other terms involving $y$.\label{sec:length1}

\subsection{The Case $D_{k+2}$}

\noindent In this section we prove the second part of {\bf Theorem 1}: \\
\noindent {\em The resolved geometry corresponding
to the Intriligator--Wecht superpotential 
$$W(x,y) = \dfrac{1}{k+1}x^{k+1}+xy^2$$
corresponds to the singular geometry
\begin{eqnarray*}
X^2 - Y^2Z + T(Z^{k/2} - T)^2 &=& 0, \skop (k \;\mathrm{even})\\
X^2 - Y^2Z - T(Z^k-T^2) &=& 0, \skop (k \;\mathrm{odd}).
\end{eqnarray*}}

\noindent \textbf{The resolved geometry} $\Hat{\MM}$\\
\noindent From the Intriligator-Wecht superpotential
$$W(x,y) = \dfrac{1}{k+1}x^{k+1} + xy^2,$$
we compute the resolved geometry $\Hat{\MM}$ in terms of transition functions
$$\beta = \gamma^{-1}, \skop v_1 = \gamma^{-1} w_1, \skop v_2 = \gamma^3 w_2 + \gamma^2 w_1^k + w_1^2.$$
To find the $\PP^1$s, we substitute $w_1(\gamma) = x + \gamma y$ into the $v_2$ transition function
$$v_2(\beta) = \beta^{-3} w_2 + \beta^{-2}(x + \beta^{-1}y)^k + (x + \beta^{-1}y)^2.$$
If we choose
$$w_2(\gamma) = \dfrac{x^k - (x+\gamma y)^k}{\gamma},$$
in the $\beta$ chart the section is
$$v_1(\beta) = \beta x + y, \skop v_2(\beta) = \beta^{-2}(x^k+y^2) + \beta^{-1}(2xy) + x^2.$$
This is only holomorphic if
$$x^k+y^2 = 2xy = 0,$$
and so we have a single $\PP^1$ located at
$$w_1(\gamma) = w_2(\gamma) = 0, \skop v_1(\beta) = v_2(\beta) = 0.$$
This is exactly what we expect from computing critical points of the superpotential
$$\d W = (x^k+y^2) dx + 2xy \;dy = 0.$$
\skp

\noindent \textbf{Global holomorphic functions}\\
\noindent The transition functions
$$\beta = \gamma^{-1}, \skop v_1 = \gamma^{-1} w_1, \skop v_2 = \gamma^3 w_2 + \gamma^2 w_1^k + w_1^2,$$
are quasi-homogeneous if we assign the weights
$$\begin{array}{c|c|c|c|c|c}
\beta & v_1 & v_2 & \gamma & w_1 & w_2 \\
k-2 & k & 4 & 2-k & 2 & 3k-2
\end{array}.$$
The global holomorphic functions will thus necessarily be quasi-homogeneous in these weights.
We find the following global holomorphic functions for $k$ even:
$$\begin{array}{c|ccccc}
4 & Z &=& v_2 &=& \gamma^3 w_2 + \gamma^2 w_1^k + w_1^2\\
2k & T &=& \beta^2 v_2 - v_1^2 &=& \gamma w_2 + w_1^k\\
3k-2 & Y &=& \beta(Z^{k/2}-T) &=& \gamma^{-1}(Z^{k/2}-T)\\
3k & X &=& v_1(Z^{k/2}-T) &=& \gamma^{-1}w_1(Z^{k/2}-T),
\end{array}$$
and a similar set of global holomorphic functions for $k$ odd:
$$\begin{array}{c|ccccc}
4 & Z &=& v_2 &=& \gamma^3 w_2 + \gamma^2 w_1^k + w_1^2\\
2k & T &=& \beta^2 v_2 - v_1^2 &=& \gamma w_2 + w_1^k\\
3k-2 & Y &=& v_1 Z^{(k-1)/2}-\beta T &=& 
\gamma^{-1}(w_1 Z^{(k-1)/2}- T)\\
3k & X &=& \beta Z^{(k+1)/2}-v_1 Y &=&
\gamma^{-1}( Z^{(k+1)/2}-w_1 T).
\end{array}$$
\skp

\noindent \textbf{The singular geometry} $\MM_0$\\
\noindent We conjecture that the ring of 
global holomorphic functions is generated by $X,Y,Z$ and $T$, 
subject to the degree $6k$ relation
$$\begin{array}{ccccccc}
\MM_0:& \;\;\; & X^2 - ZY^2 +T(Z^{k/2}-T)^2 &=& 0,& \;\;\; & k \; \mathrm{even},\\
\MM_0:& \;\;\; & X^2 - ZY^2 -T(Z^k - T^2) &=& 0,& \;\;\; & k \; \mathrm{odd}.
\end{array}$$
The functions $X,Y,Z$ and $T$ give us a blow-down map whose image $\MM_0$
has an isolated $D_{k+2}$ singularity.\skp

\noindent {\bf Review of length 2 blowup}\\
\noindent Before doing the blowup to show that we have the right blow down, we review some results from \cite[Chapter 4]{thesis}.  There we found
small resolutions of length 2 singularities by using deformations of matrix factorizations for  $D_{n+2}$ surface singularities.

The deformed $D_{n+2}$ equation was given by
$$0 = X^2 + Y^2Z - h'(ZP''^2+Q''^2)+2\delta'(YQ'' +(-1)^{m+1} XP''),$$
where for $t=0$ we have $\delta'=0$ and
$$\begin{array}{c|ccc}
m \mathrm{\;even} & h'(Z)=Z^{n-m}, & P''(Z) = i^m Z^{m/2}, & Q''(Z)=0.\\
m \mathrm{\;odd} & h'(Z)=Z^{n-m}, & P''(Z) = 0, & Q''(Z)= i^{m+1}Z^{(m+1)/2}.
\end{array}$$
The blowup was given by the equation for the Grassmannian $G(2,4)\subseteq \PP^5$
$$\alpha^2 - \varphi^2Z + (-1)^{m+1}h'\varepsilon^2 + 2i^{m+1}\delta'\varepsilon\varphi = 0,$$
in terms of Pl\"ucker coordinates
\begin{eqnarray*}
\alpha &=& iXY - i^{2m+1}h'P''Q''\\
\varepsilon &=&  i^{m+3}XP''+i^{-m+1}YQ''\\
\varphi &=& Y^2 - h'P''^2. 
\end{eqnarray*}
The interesting charts were $\varphi = 1$ and $\varepsilon = 1$, with transition functions
\begin{eqnarray}\label{eq:tf}
\varphi_2 &=& \varepsilon_1^{-1},\\
\alpha_2 &=& \varepsilon_1^{-1}\alpha_1,\\
Z &=& (-1)^{m+1} \varepsilon_1^2 h'(Z,t) + \alpha_1^2 + 2i^{m+1}\delta'\varepsilon_1,\\
t &=& t.
\end{eqnarray}
\skp

\noindent \textbf{The blowup}\\
\noindent For $k$ even, the equation for $\M_0$ corresponds to
$$\delta'=0, \skop h'=T, \skop P''=0, \skop Q''=i^k(Z^{k/2}-T), \skop n=m=k-1,$$
and Pl\"ucker coordinates
$$\alpha = iXY, \skop \varepsilon = -Y(Z^{k/2}-T), \skop \varphi = Y^2.$$
The connection with the original transition function coordinates is
$$\begin{array}{c|ccccc}
4 & Z &=& v_2 &=& \gamma^3 w_2 + \gamma^2 w_1^k + w_1^2\\
2k & T &=& \beta^2 v_2 - v_1^2 &=& \gamma w_2 + w_1^k\\
3k-2 & Y &=& \beta(Z^{k/2}-T) &=& \gamma^{-1}(Z^{k/2}-T)\\
3k & iX &=& v_1(Z^{k/2}-T) &=& \gamma^{-1}w_1(Z^{k/2}-T).
\end{array}$$
Note that
$$\begin{array}{ccc}
\beta &=& \dfrac{Y}{Z^{k/2}-T} = -\dfrac{\varphi}{\varepsilon} = -\varphi_2 \\
v_1 &=& \dfrac{iX}{Z^{k/2}-T} = -\dfrac{\alpha}{\varepsilon} = -\alpha_2\\
v_2 &=& Z \\
\gamma &=& \dfrac{Z^{k/2} - T}{Y} = -\dfrac{\varepsilon}{\varphi} = -\varepsilon_1\\
w_1 &=& \dfrac{iX}{Y} = \dfrac{\alpha}{\varphi} = \alpha_1 \\
w_2&=& \beta(T-w_1^k) 
= -\varepsilon_2^{-1}(T-\alpha_2^k)
\end{array}$$
In particular, the transition functions~\eqref{eq:tf} become
\begin{eqnarray*}
\beta &=& \gamma^{-1},\\
v_1 &=& \gamma^{-1}w_1,\\
v_2 &=& \gamma^2 T + w_1^2 = \gamma^3 w_2 + \gamma^2 w_1^k + w_1^2.
\end{eqnarray*}

The blowup for $k$ odd is similar.  For more details about the Grassmann blowup for singularities of type $D_n$ see \cite[pp. 21-23]{flops}.

\label{sec:length2}

\subsection{The Case $E_7$}

\noindent In this section we prove the third part of {\bf Theorem 1}: \\
\noindent {\em The resolved geometry corresponding
to the Intriligator--Wecht superpotential 
$$W(x,y) = \dfrac{1}{3}y^3+yx^3$$
corresponds to the singular geometry
$$ X^2 - Y^3 + Z^5 + 3TYZ^2 + T^3Z = 0. $$}

\noindent \textbf{The resolved geometry} $\Hat{\MM}$\\
From the Intriligator-Wecht superpotential
$$W(x,y) = \dfrac{1}{3}y^3 + yx^3,$$
we compute the resolved geometry $\Hat{\MM}$ in terms of transition functions
$$\beta = \gamma^{-1}, \skop v_1 = \gamma^{-1} w_1, \skop v_2 = \gamma^3 w_2 + \gamma w_1^3 + 
\gamma^{-1} w_1^2.$$
To find the $\PP^1$s, we substitute $w_1(\gamma) = x + \gamma y$ into the $v_2$ transition function
$$v_2(\beta) = \beta^{-3} w_2 + \beta^{-1}(x + \beta^{-1}y)^3 + \beta(x + \beta^{-1}y)^2.$$
If we choose
$$w_2(\gamma) = \dfrac{x^3 + 3\gamma x^2y - (x+\gamma y)^3}{\gamma^2},$$
in the $\beta$ chart the section is
$$v_1(\beta) = \beta x + y, \skop v_2(\beta) = \beta^{-2}(3x^2 y) + \beta^{-1}(x^3 + y^2)+2xy+\beta x^2.$$
This is only holomorphic if
$$3x^2y = x^3+y^2 = 0,$$
and so we have a single $\PP^1$ located at
$$w_1(\gamma) = w_2(\gamma) = 0, \skop v_1(\beta) = v_2(\beta) = 0.$$
This is exactly what we expect from computing critical points of the superpotential
$$\d W = 3x^2y\; dx + (y^2+x^3)dy = 0.$$
\skp

\noindent \textbf{Global holomorphic functions}\\
\noindent The transition functions
$$\beta = \gamma^{-1}, \skop v_1 = \gamma^{-1} w_1, \skop v_2 = \gamma^3 w_2 + \gamma w_1^3 +
\gamma^{-1} w_1^2,$$
are quasi-homogeneous if we assign the weights
$$\begin{array}{c|c|c|c|c|c}
\beta & v_1 & v_2 & \gamma & w_1 & w_2 \\
1 & 3 & 5 & -1 & 2 & 8
\end{array}.$$
The global holomorphic functions will thus necessarily be quasi-homogeneous in these weights.
We find the following global holomorphic functions:
$$\begin{array}{c|ccccc}
6 & X &=& \beta v_2-v_1^2 && \\
8 & Y &=& v_1v_2 - \beta^2X && \\
10 & Z &=& v_2^2-\beta v_1 X && \\
15 & F &=& v_2^3 - 2v_1^3X + (\beta^3-3v_1)X^2. && 
\end{array}$$
\skp

\noindent \textbf{The singular geometry} $\MM_0$\\
We conjecture that the ring of 
global holomorphic functions is generated by $X,Y,Z$ and $F$, 
subject to the degree $30$ relation
$$\MM_0: \;\;\; F^2 -Z^3 + X^5 + 3X^2YZ + XY^3 = 0.$$
Do the functions $X,Y,Z$ and $F$ give us a blow-down map whose image $\MM_0$
has an isolated $E_7$ singularity?\skp

\noindent \textbf{The blowup}\\
\noindent We now verify that we have identified the right singular space $\MM_0$ by inverting
the blow-down.  In the $\beta$ and $\gamma$ charts we find
$$\begin{array}{ccc|cccc}
\beta &=& \dfrac{Y^2+XZ}{F} & \gamma &=& \dfrac{F}{Y^2+XZ}&\\
&&&&&&\\
v_1 &=& \dfrac{X^3 + YZ}{F} & w_1 &=& \dfrac{X^3 + YZ}{Y^2 + XZ}&\\
&&&&&&\\
v_2 &=& \dfrac{Z^2 - X^2Y}{F} & w_2 &=& \dfrac{X^4 - Y^3}{Y^2 + XZ}& = w_1X-Y.
\end{array}$$
This suggests that we should rewrite the equation for $\MM_0$ as
$$\MM_0: \;\;\; F^2  + XY(Y^2 + XZ) + X^2(X^3+YZ) - Z(Z^2 - X^2Y) = 0,$$ 
and that we can obtain $\Hat{\MM}$ by blowing up
$$F = Y^2+XZ = X^3+YZ = Z^2-X^2Y = 0.$$

\noindent \textbf{The locus} $\C$\\
\noindent Let $\S$ denote the surface
$$\S:\;\; F=Y^2+XZ=0.$$
Our $\Hat{M}$ coordinate patches $(\beta,v_1,v_2)$ and $(\gamma,w_1,w_2)$ cover
everything except the locus 
$$\C = \S \cap \MM_0.$$
The intersection of $\S$ with the 3-fold $\MM_0$ yields the new equation
$$X^5 + 2X^2YZ - Z^3 = 0.$$
(This was obtained by finding the Groebner basis for the ideal generated by $F$, $Y^2+XZ$, and
the equation for $\MM_0$.)  
\skp

\noindent For $X \neq 0$ we can write
$$Z = -\dfrac{Y^2}{X},$$
and so the equations for $\C$ become
$$\C: \;\;F = Y^2 + XZ = (X^4 - Y^3)^2 = 0, \skop (X \neq 0).$$
We can parametrize this curve by
\begin{eqnarray*}
X &=& t^3,\\
Y &=& t^4,\\
Z &=& -t^5,\\
F &=& 0.
\end{eqnarray*}
From this we see that blowing up $\C \subset \MM_0$ is equivalent to blowing up
$$F = Y^2+XZ = X^3+YZ = Z^2-X^2Y = X^4 - Y^3 = 0.$$
In this case we would have additional coordinates $v_3, w_3$ for the blowup:
$$\begin{array}{ccc|cccc}
\beta &=& \dfrac{Y^2+XZ}{F} & \gamma &=& \dfrac{F}{Y^2+XZ}&\\
&&&&&&\\
v_1 &=& \dfrac{X^3 + YZ}{F} & w_1 &=& \dfrac{X^3 + YZ}{Y^2 + XZ}&\\
&&&&&&\\
v_2 &=& \dfrac{Z^2 - X^2Y}{F} & w_2 &=& \dfrac{X^4 - Y^3}{Y^2 + XZ}&\\
&&&&&&\\
v_3 &=& \dfrac{X^4-Y^3}{F} & w_3 &=& \dfrac{Z^2 - X^2Y}{Y^2+XZ}.&
\end{array}$$
Note that both $v_3$ and $w_3$ add nothing new, as we can solve for them
in terms of the other coordinates:
$$\begin{array}{ccccc}
v_3 &=& \gamma^{-1}w_2 &=& \beta^4 v_2 - \beta^3 v_1^2 - v_1^3,\\
w_3 &=& \beta^{-1} v_2 &=& \gamma^4 w_2 + \gamma^2 w_1^3 + w_1^2.
\end{array}$$
These are precisely the additional coordinates we introduced in our resolution of the ideal sheaf (see Section 3.5).
Finally, with these identifications we find transition functions
$$\beta = \gamma^{-1}, \skop v_1 = \gamma^{-1} w_1, \skop v_2 = \gamma^3 w_2 + \gamma w_1^3 + 
\gamma^{-1} w_1^2,$$
which are exactly the ones we started with.

\label{sec:length3}

\section{Proof of Theorem 2}\label{sec:thm2}

In this section we analyze the extra `hat' cases in 
the Intriligator--Wecht classification of superpotentials.  In particular, we prove
Theorem 2 by finding singular geometries corresponding to each of the 
$\Hat{O},\Hat{A},\Hat{D}$ and $\Hat{E}$ cases.
As in the case of Theorem 1, we first find global holomorphic functions using the algorithm described in the Appendix,
and then verify the resulting singular space by blowing back up to recover our original transition functions.

\subsection{The Case $\Hat{O}$}

\noindent \textbf{The resolved geometry} $\Hat{\MM}$\\
\noindent From the Intriligator-Wecht superpotential
$$W(x,y) = 0,$$
we compute the resolved geometry $\Hat{\MM}$ in terms of transition functions
$$\beta = \gamma^{-1}, \skop v_1 = \gamma^{-1} w_1, \skop v_2 = \gamma^3 w_2.$$
To find the $\PP^1$s, we substitute $w_1(\gamma) = x + \gamma y$ into the $v_2$ transition function
$$v_2(\beta) = \beta^{-3} w_2.$$
If we choose
$$w_2(\gamma) = 0,$$
in the $\beta$ chart the section is
$$v_1(\beta) = \beta x + y, \skop v_2(\beta) = 0.$$
This is holomorphic for all $x$ and $y$,
and so we have a 2-parameter family of $\PP^1$s located at
$$w_1(\gamma) = x+\gamma y, w_2(\gamma) = 0, \skop v_1(\beta) = \beta x + y, v_2(\beta) = 0.$$
This is exactly what we expect from computing critical points of the superpotential
$$\d W = 0.$$
\skp

\noindent \textbf{Global holomorphic functions}\\
\noindent The transition functions
$$\beta = \gamma^{-1}, \skop v_1 = \gamma^{-1} w_1, \skop v_2 = \gamma^3 w_2$$
are quasi-homogeneous if we assign the weights
$$\begin{array}{c|c|c|c|c|c}
\beta & v_1 & v_2 & \gamma & w_1 & w_2 \\
1 & d+1 & e-3 & -1 & d & e
\end{array}.$$
Notice the freedom in choosing $d$ and $e$: there is a two-dimensional lattice of possible weight 
assignments. The global holomorphic functions will necessarily be quasi-homogeneous in these weights.
We find global holomorphic functions:
$$X_{ij} = \beta^{i} v_1^j v_2 = \gamma^{3-i-j} w_1^j w_2, \skop i,j \geq 0 ,\;\;\; i+j \leq 3.$$
\skp

\noindent \textbf{The singular geometry} $\MM_0$\\
\noindent If we rewrite our functions in a homogeneous manner as
$$\til{X}_{ij} = a^{3-i-j} b^i c^j, \skop i,j \geq 0 ,\;\;\; i+j \leq 3,$$
we can now identify the ring of global holomorphic functions as homogeneous
polynomials of degree 3 in 3 variables.  In other words, the ring is isomorphic to
$$\CC[a,b,c]^{\ZZ_3},$$
and our singular variety is simply
$$\MM_0: \;\;\; \CC^3/\ZZ_3.$$
\skp

\noindent \textbf{The blowup}\\
\noindent We now verify that we have identified the right singular space $\MM_0$ by inverting
the blow-down.  In the $\beta$ and $\gamma$ charts we find
$$\begin{array}{ccc|cccc}
\beta &=& X_{10}/X_{00} = \til{X}_{10}/\til{X}_{00} = b/a & \gamma &=& a/b &\\
&&&&&&\\
v_1 &=& X_{01}/X_{00} = \til{X}_{01}/\til{X}_{00} = c/a & w_1 &=& c/b&\\
&&&&&&\\
v_2 &=& X_{00} = \til{X}_{00} = a^3 & w_2 &=& b^3 &
\end{array}$$
which gives transition functions
$$\beta = \gamma^{-1}, \skop v_1 = \gamma^{-1} w_1, \skop v_2 = \gamma^3 w_2.$$
These are precisely what we started with!\skp

\noindent \textbf{Remark}\\
\noindent In the resolved $\Hat{\M}$ geometry, what we have is a $\PP^1$ inside a $\PP^2$ (or any other del Pezzo surface).  If you have a $\PP^2$ inside a Calabi-Yau and blow it down, you get $\CC^3/\ZZ^3$ as the singular point.

\subsection{The Case $\Hat{A}$}

\noindent \textbf{The resolved geometry} $\Hat{\MM}$\\
\noindent From the Intriligator-Wecht superpotential
$$W(x,y) = \dfrac{1}{2}x^2,$$
we compute the resolved geometry $\Hat{\MM}$ in terms of transition functions
$$\beta = \gamma^{-1}, \skop v_1 = \gamma^{-1} w_1, \skop v_2 = \gamma^3 w_2 + \gamma^2 w_1.$$
To find the $\PP^1$s, we substitute $w_1(\gamma) = x + \gamma y$ into the $v_2$ transition function
$$v_2(\beta) = \beta^{-3} (w_2 + y) + \beta^{-2}x.$$
If we choose
$$w_2(\gamma) = -y,$$
in the $\beta$ chart the section is
$$v_1(\beta) = \beta x + y, \skop v_2(\beta) = \beta^{-2}x.$$
This is only holomorphic $x=0$. Since $y$ is free,
we have a 1-parameter family of $\PP^1$s located at
$$w_1(\gamma) = \gamma y, w_2(\gamma) = -y, \skop v_1(\beta) = y, v_2(\beta) = 0.$$
This is exactly what we expect from computing critical points of the superpotential
$$\d W = x\d x = 0.$$
\skp

\noindent \textbf{Global holomorphic functions}\\
\noindent The transition functions
$$\beta = \gamma^{-1}, \skop v_1 = \gamma^{-1} w_1, \skop v_2 = \gamma^3 w_2 + \gamma^2 w_1$$
are quasi-homogeneous if we assign the weights
$$\begin{array}{c|c|c|c|c|c}
\beta & v_1 & v_2 & \gamma & w_1 & w_2 \\
1 & d+1 & d-2 & -1 & d & d+1
\end{array}.$$
Notice the freedom in choosing $d$: there is a one-dimensional lattice of possible weight 
assignments. The global holomorphic functions will necessarily be quasi-homogeneous in these weights.
We find global holomorphic functions:
$$\begin{array}{c|ccc}
d-2 & y_1 &=& v_2 = \gamma^3 w_2 + \gamma^2 w_1\\
d-1 & y_2 &=& \beta v_2 = \gamma^2 w_2 + \gamma w_1\\
d   & y_3 &=& \beta^2 v_2 = \gamma w_2 + w_1\\
d+1 & y_4 &=& \beta^3 v_2 - v_1 = w_2
\end{array}$$
\skp

\noindent \textbf{The singular geometry} $\MM_0$\\
\noindent The functions $y_i$ satisfy the single degree $2d-2$ relation
$$y_2^2 - y_1 y_3 = 0,$$
with $y_4$ free.  In other words, our singular geometry $\MM_0$ is a curve
of $A_1$ singularities, parametrized by $y_4$.
\skp

\noindent\textbf{The blowup}\\
\noindent We now verify that we have identified the right singular space $\MM_0$ by inverting
the blow-down.  In the $\beta$ and $\gamma$ charts we find
$$\begin{array}{ccc|cccc}
\beta &=& y_2/y_1 & \gamma &=& y_1/y_2 &\\
&&&&&&\\
v_1 &=& (y_2y_3-y_1y_4)/y_1 & w_1 &=& (y_2y_3-y_1y_4)/y_2&\\
&&&&&&\\
v_2 &=& y_1 & w_2 &=& y_4. &
\end{array}$$
This gives transition functions
$$\beta = \gamma^{-1}, \skop v_1 = \gamma^{-1} w_1, \skop v_2 = \gamma^3 w_2 + \gamma^2 w_1,$$
as expected.


\subsection{The Case $\Hat{D}$}

\noindent \textbf{The resolved geometry} $\Hat{\MM}$\\
\noindent From the Intriligator-Wecht superpotential
$$W(x,y) = x^2 y,$$
we compute the resolved geometry $\Hat{\MM}$ in terms of transition functions
$$\beta = \gamma^{-1}, \skop v_1 = \gamma^{-1} w_1, \skop v_2 = \gamma^3 w_2 + \gamma w_1^2.$$
To find the $\PP^1$s, we substitute $w_1(\gamma) = x + \gamma y$ into the $v_2$ transition function
$$v_2(\beta) = \beta^{-3}(w_2+y^2) + \beta^{-2}(2xy) + \beta^{-1}x^2.$$
If we choose
$$w_2(\gamma) = -y^2,$$
in the $\beta$ chart the section is
$$v_1(\beta) = \beta x + y, \skop v_2(\beta) = \beta^{-2}(2xy) + \beta^{-1}x^2.$$
This is only holomorphic if $x=0$. Since $y$ is free,
we have a 1-parameter family of $\PP^1$s located at
$$w_1(\gamma) = \gamma y, w_2(\gamma) = -y^2, \skop v_1(\beta) = y, v_2(\beta) = 0.$$
This is exactly what we expect from computing critical points of the superpotential
$$\d W = 2xy \;\d x + x^2\d y = 0.$$
\skp

\noindent \textbf{Global holomorphic functions}\\
\noindent The transition functions
$$\beta = \gamma^{-1}, \skop v_1 = \gamma^{-1} w_1, \skop v_2 = \gamma^3 w_2 + \gamma w_1^2$$
are quasi-homogeneous if we assign the weights
$$\begin{array}{c|c|c|c|c|c}
\beta & v_1 & v_2 & \gamma & w_1 & w_2 \\
1 & d+1 & 2d-1 & -1 & d & 2d+2
\end{array}.$$
Notice the freedom in choosing $d$: there is a one-dimensional lattice of possible weight 
assignments. The global holomorphic functions will necessarily be quasi-homogeneous in these weights.
We find global holomorphic functions:
$$\begin{array}{c|ccc}
2d-1 & y_1 &=& v_2 = \gamma^3 w_2 + \gamma w_1^2\\
2d & y_2 &=& \beta v_2 = \gamma^2 w_2 + w_1^2\\
3d   & y_3 &=& v_1 v_2 = \gamma^2 w_1 w_2 + w_1^3\\
2d+2 & y_4 &=& \beta^3 v_2 - v_1^2 = w_2
\end{array}$$
\skp

\noindent \textbf{The singular geometry} $\MM_0$\\
\noindent The functions $y_i$ satisfy the single degree $6d$ relation
$$y_3^2 - y_2^3 + y_1^2 y_4 =0.$$
\skp

\noindent \textbf{The blowup}\\
We now verify that we have identified the right singular space $\MM_0$ by inverting
the blow-down.  In the $\beta$ and $\gamma$ charts we find
$$\begin{array}{ccc|cccc}
\beta &=& y_2/y_1 & \gamma &=& y_1/y_2 &\\
&&&&&&\\
v_1 &=& y_3/y_1 & w_1 &=& y_3/y_2&\\
&&&&&&\\
v_2 &=& y_1 & w_2 &=& y_4, &
\end{array}$$
with transition functions
$$\beta = \gamma^{-1}, \skop v_1 = \gamma^{-1} w_1, \skop v_2 = \gamma^3 w_2 + \gamma w_1^2.$$

\subsection{The Case $\Hat{E}$}

\noindent \textbf{The resolved geometry} $\Hat{\MM}$\\
\noindent From the Intriligator-Wecht superpotential
$$W(x,y) = \dfrac{1}{3}x^3,$$
we compute the resolved geometry $\Hat{\MM}$ in terms of transition functions
$$\beta = \gamma^{-1}, \skop v_1 = \gamma^{-1} w_1, \skop v_2 = \gamma^3 w_2 + \gamma^2 w_1^2.$$
To find the $\PP^1$s, we substitute $w_1(\gamma) = x + \gamma y$ into the $v_2$ transition function
$$v_2(\beta) = \beta^{-3}(w_2+2xy+\gamma y^2) + \beta^{-2}x^2.$$
If we choose
$$w_2(\gamma) = -2xy-\gamma y^2,$$
in the $\beta$ chart the section is
$$v_1(\beta) = \beta x + y, \skop v_2(\beta) = \beta^{-2}x^2.$$
This is only holomorphic if $x=0$. Since $y$ is free,
we have a 1-parameter family of $\PP^1$s located at
$$w_1(\gamma) = \gamma y, w_2(\gamma) = -\gamma y^2, \skop v_1(\beta) = y, v_2(\beta) = 0.$$
This is exactly what we expect from computing critical points of the superpotential
$$\d W = x^2 \;\d x = 0.$$
\skp

\noindent \textbf{Global holomorphic functions}\\
\noindent The transition functions
$$\beta = \gamma^{-1}, \skop v_1 = \gamma^{-1} w_1, \skop v_2 = \gamma^3 w_2 + \gamma^2 w_1^2$$
are quasi-homogeneous if we assign the weights
$$\begin{array}{c|c|c|c|c|c}
\beta & v_1 & v_2 & \gamma & w_1 & w_2 \\
1 & d+1 & 2d-2 & -1 & d & 2d+1
\end{array}.$$
Notice the freedom in choosing $d$: there is a one-dimensional lattice of possible weight 
assignments. The global holomorphic functions will necessarily be quasi-homogeneous in these weights.
We find global holomorphic functions:
$$\begin{array}{c|ccc}
2d-2 & y_1 &=& v_2 \\
2d-1 & y_2 &=& \beta v_2 \\
2d   & y_3 &=& \beta^2 v_2 \\
3d-1 & y_4 &=& v_1 v_2\\
3d   & y_5 &=& \beta v_1 v_2\\
4d   & y_6 &=& v_1^2 v_2\\
4d+1 & y_7 &=& \beta(\beta^4 v_2-v_1^2)v_2 = \beta v_3 v_2\\
5d+1 & y_8 &=& v_1(\beta^4 v_2-v_1^2)v_2 = v_1 v_3 v_2\\
6d+2 & y_9 &=& (\beta^4 v_2-v_1^2)^2 v_2 = v_3^2 v_2
\end{array}$$
where we have defined
$$v_3 = \beta^4 v_2-v_1^2 = \gamma^{-1}w_2.$$
\skp

\noindent \textbf{The singular geometry} $\MM_0$\\
\noindent The functions $y_i$ satisfy a total of 20 distinct relations, most of which are obvious.
To simplify things, consider the monomial mapping
$$\beta^i v_1^j v_3^k v_2 \longmapsto a^{2-i-j-k} b^i c^j f^k.$$
Our functions now become
$$\begin{array}{c|ccc}
2d-2 & y_1 &=& a^2 \\
2d-1 & y_2 &=& ab \\
2d   & y_3 &=& b^2 \\
3d-1 & y_4 &=& ac\\
3d   & y_5 &=& bc\\
4d   & y_6 &=& c^2\\
4d+1 & y_7 &=& bf\\
5d+1 & y_8 &=& cf\\
6d+2 & y_9 &=& f^2
\end{array}$$
Note that
$$\begin{array}{ccccc}
\beta &=& y_2/y_1 &=& b/a\\
v_1 &=& y_4/y_1 &=& c/a\\
v_2 &=& y_1 &=& a^2\\
v_3 &=& y_7/y_2 &=& f/a,
\end{array}$$
so the relation defining $v_3$ becomes
$$af = b^4 - c^2.$$
This means we can add the function $af$ to our list, together with the relation:
$$y_{10} = af = b^4 - c^2.$$
\skp

\noindent Now the functions $y_1,...,y_{10}$ are exactly the 10 monomials of degree 2 in 4 variables, together
with the above relation.  The ring of global holomorphic functions is thus
$$(\CC[a,b,c,f]/\ZZ_2)/(af-b^4+c^2),$$
where the $\ZZ_2$ acts diagonally as -1.  In other words, we have a hypersurface in a $\ZZ_2$
quotient space:
$$(b^2+c)(b^2-c)=af \skop \mathrm{in} \;\;\;\CC^4/\ZZ_2.$$
We can immediately see from this equation that a small resolution, where we blow up an ideal of
the form
$$b^2+c=a=0,$$
won't work, since the $\ZZ_2$ action interchanges $b^2+c$ and $b^2-c$.  We will need to do
a big blow up of the origin instead.
\skp

\noindent \textbf{The blowup}\\
\noindent We now verify that we have identified the right singular space $\MM_0$ by inverting
the blow-down.  In the $\beta$ and $\gamma$ charts we find
$$\begin{array}{ccc|cccc}
\beta &=& b/a & \gamma &=& a/b &\\
&&&&&&\\
v_1 &=& c/a & w_1 &=& c/b &\\
&&&&&&\\
v_2 &=& a^2 & w_2 &=& f/b &\\
&&&&&&\\
v_3 &=& f/a & w_3 &=& b^2 &
\end{array}$$
We will perform the big blowup of the origin, with corresponding $\PP^3$ coordinates:
$$\begin{array}{ccccccccc}
a &=& b &=& c &=& f &=& 0.\\
\alpha && \delta && \rho && \nu &&
\end{array}$$
Note that all eight coordinates switch sign under the $\ZZ_2$ action.\skp
The blowup has four coordinate charts
$$\begin{array}{c|c|c|c}
\alpha = 1 & \delta = 1 & \rho = 1 & \nu = 1\\
&&&\\
a=a & a = \alpha_2 b & a = \alpha_3 c & a = \alpha_4 f\\
b=\delta_1 a & b=b & b=\delta_3 c & b = \delta_4 f\\
c = \rho_1 a & c=\rho_2 b & c=c & c = \rho_4 f\\
f = \nu_1 a & f = \nu_2 b & f = \nu_3 c & f=f\\
&&&\\
\nu_1 = \delta_1^4 a^2-\rho_1^2 & b^2 = \alpha_2\nu_2+\rho_2^2 & \alpha_3\nu_3 = \delta_3^4c^2-1 &
\alpha_4 = \delta_4^4 f^2 - \rho_4^2\\
&&&\\
(a^2,\delta_1,\rho_1) & (\alpha_2,\rho_2,\nu_2) & (\alpha_3,\delta_3,c^2,\nu_3) & (\delta_4,\rho_4,f^2)  
\end{array}$$
\skp

\noindent \textbf{Remarks}
\begin{itemize}
\item The functions $\alpha_i,\delta_i,\rho_i,$ and $\nu_i$ are all invariant under the $\ZZ_2$ action,
since they are all ratios of functions which change sign:
$$\delta_1 = \delta/\alpha, \;\;\rho_2 = \rho/\delta, \;\;...\; \mathrm{etc.}$$
\item Because $a,b,c,$ and $f$ all change sign under the $\ZZ_2$ action, we must take their invariant
counterparts $a^2, b^2, c^2,$ and $f^2$ when we list the final coordinates for each chart.
\item In the $\delta=1$ chart, we solve for $b^2$ instead of $b$, because $b$ is not an invariant function.
\item The blow up is nonsingular.  In the $\alpha = 1, \delta = 1,$ and $\nu=1$ charts we see this
because we are left with three coordinates and no relations, so these charts are all isomorphic to $\CC^3$.
In the $\rho = 1$ chart, we have a hypersurface in $\CC^4$ defined by the non-singular equation
$$\alpha_3 \nu_3 = \delta_3^4 c^2 -1.$$
\end{itemize}

\noindent \textbf{Transition functions}\\
\noindent Between the first two charts $\alpha = 1$ and $\delta = 1$, we have transition functions
\begin{eqnarray*}
\delta_1 &=& \delta/\alpha = \alpha_2^{-1}\\
\rho_1 &=& \rho/\alpha = (\delta/\alpha)(\rho/\delta) = \alpha_2^{-1} \rho_2\\
a^2 &=& \alpha_2^2 b^2 = \alpha_2^2 (\alpha_2 \nu_2 + \rho_2^2)\\
&=& \alpha_2^3\nu_2 + \alpha_2^2\rho_2^2
\end{eqnarray*}
Notice that
$$\begin{array}{ccccc|ccccc}
\delta_1 &=& b/a &=& \beta & \alpha_2 &=& a/b &=& \gamma\\
\rho_1 &=& c/a &=& v_1 & \rho_2 &=& c/b &=& w_1\\
a^2 &=& a^2 &=& v_2 & \nu_2 &=& f/b &=& w_2,
\end{array}$$
and so our transition functions are really
$$\beta = \gamma^{-1}, \skop v_1 = \gamma^{-1} w_1, \skop v_2 = \gamma^3 w_2 + \gamma^2 w_1^2.$$
These are exactly the $\Hat{E}$ transition functions we started with!

\subsection{Comparison with ADE cases}

We have seen that the singular geometries corresponding to Intriligator and Wecht's `hat' cases are given
by

$$\begin{array}{c|c|c}
\Hat{O} & W(x,y) = 0 & \CC^3/\ZZ_3\\
&&\\
\Hat{A} & W(x,y) = \dfrac{1}{2}y^2 & \CC[X,Y,Z,T]/(XY-Z^2) \cong \CC \times \CC^2/\ZZ_2\\
& & \mathrm{geometry\;has\;curve\;of\;} A_1 \mathrm{singularities}\\
&&\\
\Hat{D} & W(x,y) = xy^2 & y_1^2 - y_2^3 + y_3^2 y_4 = 0\\
&&\\
& & \mathrm{recall\;geometry\;for\;} D_{k+2}:\\
&& y_1^2 + y_2^3 + y_3^2y_4 + y_4^k y_2 = 0\\
&&\\
\Hat{E} & W(x,y) = \dfrac{1}{3}y^3 & (\CC[a,b,u,v]/\ZZ_2)/(b^4 - u^2 - av)\\
& & \mathrm{This\;is\;a\;hypersurface\;in\;} \CC^4/\ZZ_2.
\end{array}$$

Note that in both the $\Hat{A}$ and $\Hat{D}$ cases, the resulting equations can be obtained from the $A_k$ and
$D_{k+2}$ equations by dropping the $k$-dependent terms.  In other words, we are tempted to think of $\Hat{A}$ and $\Hat{D}$ as the $k \rightarrow \infty$ limit.  Perhaps in trying to come up with an analogous statement for $\Hat{E}$ we can learn something about the ``missing'' $E_6$ and $E_8$ cases.  In particular, it will be interesting to understand the role of these spaces in a geometric model for RG flow.

\section{Conclusions}

We end by posing a series of questions for the future which are beyond the scope of this work.

From the Intriligator--Wecht classification, we still need to understand the $E_6$ and $E_8$ cases.  
Using our algorithm we have found many global holomorphic functions, but not enough to 
give us the blow-down \cite{thesis}.  There are also questions which arise from the extra `hat' cases. 
What is the interpretation of the $\Hat{O},\Hat{A},\Hat{D},$ and $\Hat{E}$ geometries from the string
theory perspective?  The $\PP^1$s are no longer isolated; do they correspond to D-branes wrapping families of $\PP^1$s?  Moreover, what is the role of higher order terms in the superpotential?  

Do we have a geometric model for RG flow?  Proposition 1 suggests that the geometry might encode something about the RG fixed points of the corresponding matrix models or gauge theories.  Can Proposition 1 be extended?  Finding more general coordinate changes which can show how the rest of the terms
in the superpotential are affected when a bundle-changing coordinate is ``integrated out'' is a necessary step in
developing this kind of geometric picture.

Furthermore, Intriligator and Wecht have a chart of all possible flows between the RG fixed points.  We can make a similar chart based on our geometric framework.  Do they match?
Finally, what is the role of fundamentals?   Our entire analysis involves only adjoint fields, which correspond geometrically to parameters of the $\PP^1$ deformation space.  Intriligator and Wecht only find the ADE classification for superpotentials involving 2 adjoint fields, but their paper also analyzes many cases with fundamentals.  Is it possible to have a geometric interpretation for these fields?

As far as Ferrari's construction is concerned, there are many open ends to be explored.  Can we generalize Ferrari's framework to include perturbation terms for both $v_1$ and $v_2$ transition functions?  Can we generalize for cases where the geometry is specified by more than two charts?  This would enable more flexibility in 
identifying superpotentials in a ``bottom-up'' approach.  On the other hand, the techniques developed in \cite{aspinwall} in principle allow computation of the superpotential in general.  In cases where the superpotential cannot be easily identified in the transition functions, perhaps this approach should be used instead.

Moreover, in all of our new cases there is still
work to be done to complete the remaining steps in Ferrari's program.  For example, what is the solution to the matrix model corresponding to the length 3 singularity?  And what can we learn about the matrix models corresponding to the `hat' cases?  Although the singularities are no longer isolated, is it still possible to compute resolvents from the geometry?  If Ferrari's conjecture about the Calabi-Yau geometry encoding the solution to the matrix model is correct, we should now be able to solve the matrix models corresponding to the length 3 and `hat' cases.  If solutions are already known (or can be computed using traditional matrix model techniques), these examples will provide new tests to the conjecture.

\section{Appendix}

We reformulate our problem of finding global holomorphic functions as an
ideal membership problem.  We begin by illustrating the reformulation in an example.
Consider the resolved geometry for the $E_7$ Intriligator-Wecht potential:
$$\beta = \gamma^{-1}, \skop v_1 = \gamma^{-1} w_1, \skop
v_2 = \gamma^3 w_2 + \gamma w_1^3 + \gamma^{-1} w_1^2.$$
We can think of this as describing a variety in $\CC^6$, defined
by the following ideal $I \subset \CC[\gamma,w_1,w_2,\beta,v_1,v_2]$:
$$I = \langle \beta\gamma-1, v_1-\beta w_1, v_2-\gamma^3 w_2 - \gamma w_1^3 - \beta w_1^2 \rangle.$$
In order to blow down the exceptional $\PP^1$, we must find functions which are holomorphic
in each coordinate chart, and will therefore be constant on the $\PP^1$.  Such global holomorphic
functions correspond to elements of the ideal $I$ that can be written in the form\footnote{Note that in our
example, none of the defining generators for $I$ are of this form!}
$$f - g \in I, \:\;\; \where \;\;\; f \in \CC[\beta, v_1, v_2], \;\; g \in \CC[\gamma, w_1, w_2].$$
For each such element, the global holomorphic function is $f = g$.

In general, we begin with transition functions
$$\beta = \gamma^{-1}, \skop v_1=\gamma^{-n}w_1, \skop
v_2=\gamma^{-m}w_2+\partial_{w_1}E(\gamma,w_1),$$
and form the ideal
$$I = \langle \beta\gamma-1, v_1-\beta^n w_1, v_2-\beta^m w_2 - \partial_{w_1}\til{E}(\gamma,w_1)\rangle,$$
where $\til{E}(\gamma,w_1)$ is obtained from $E(\gamma,w_1)$ by replacing all instances of $\gamma^{-1}$ 
with $\beta$.

\subsection{The algorithm}
\def\pure{\mathrm{pure}}
\def\mixed{\mathrm{mixed}}

Consider a monomial $\beta^i v_1^j v_2^k \in \CC[\beta, v_1, v_2]$.  Using Groebner basis techniques,
we can easily reduce this modulo the ideal
$$I = \langle \beta\gamma-1, v_1-\beta^n w_1, v_2-\beta^m w_2 - \partial_{w_1}\til{E}(\gamma,w_1)\rangle,$$
which is determined by our particular geometry.  In general, we will find
$$\beta^i v_1^j v_2^k \stackrel{\mathrm{mod\;}I}{\equiv} \pure(\beta,v_1,v_2) + 
\pure (\gamma,w_1,w_2) + \mixed, $$
where ``pure'' and ``mixed'' stand for pure and mixed terms\footnote{We will refer to any monomial in $\CC[\gamma,w_1,w_2,\beta,v_1,v_2]$ which does not belong to either $\CC[\beta,v_1,v_2]$ or $\CC[\gamma,w_1,w_2]$ as a mixed term.} in the appropriate variables.  We can then
bring the $\pure(\beta,v_1,v_2)$ terms to the left hand side, ``updating'' our initial monomial to
the polynomial
$$\beta^i v_1^j v_2^k -\pure(\beta,v_1,v_2)\stackrel{\mathrm{mod\;}I}{\equiv} 
\pure (\gamma,w_1,w_2) + \mixed. $$
Now the challenge is to find a linear combination $f$ of such polynomials in $\CC[\beta,v_1,v_2]$ such that
the mixed terms cancel, and we are left with 
$$ f \stackrel{\mathrm{mod\;}I}{\equiv} g, \:\;\; \where \;\;\; f \in \CC[\beta, v_1, v_2], \;\; g \in \CC[\gamma, w_1, w_2].$$

The central idea (as in the Euclidean division algorithm) is to put a term order on the mixed terms we are
trying to cancel.  In this way, we can make sure we are cancelling mixed terms in an efficient manner, and the 
cancellation procedure terminates.  Because mixed terms (such as $\beta w_1$) correspond to ``poles'' in the $\gamma$ coordinate chart (such as $\gamma^{-1} w_1$), we use the weighted degree term order
\begin{mapleinput}
\mapleinline{active}{1d}{TP:=wdeg([1,1,1,-1,0,0],[b,v[1],v[2],g,w[1],w[2]]):}{}
\end{mapleinput}
\noindent which keeps track of the degree of the poles in $\gamma$.

Beginning with the superpotential, our algorithm thus consists of the following steps:
\begin{enumerate}
\item Compute transition functions following Ferrari's framework.  This gives an ideal 
$$I = \langle \beta\gamma-1, v_1-\beta^n w_1, v_2-\beta^m w_2 - \partial_{w_1}\til{E}(\gamma,w_1)\rangle \subset \CC[\gamma,w_1,w_2,\beta,v_1,v_2].$$
\item Find a Groebner basis $G$ for the ideal $I$, with respect to a term order $T$.
\item Generate a list $L$ of monomials in $\beta, v_1,$ and $v_2$ (up to some degree).
\item Reduce monomial $L[j]$ mod $I$, using $G$. What you have is
$$L[j] = \beta^i v_1^j v_2^k \stackrel{\mathrm{mod\;}I}{\equiv} \pure(\beta,v_1,v_2) + 
\pure (\gamma,w_1,w_2) + \mixed $$
where ``pure'' and ``mixed'' stand for pure and mixed terms in the appropriate variables.
Bring the $\pure(\beta,v_1,v_2)$ terms over to the LHS to make a polynomial
$$\beta^i v_1^j v_2^k - \pure(\beta,v_1,v_2) \in \CC[\beta,v_1,v_2].$$
      \begin{itemize}
      \item Record this polynomial in the array $F$ as $F[j,1]$.
      \item Record the leading term (with respect to the term order TP) of the ``mixed'' part as $F[j,2]$, and store the leading coefficient as $F[j,3]$.
      \end{itemize}
\item {Reduction routine}
      \begin{itemize}
      \item Cycle through the list of previous polynomials $F[1..j-1,*]$ and cancel leading mixed terms as much as possible.
      \item The result is a new ``updated'' polynomial $F[j,1]$ which is {\em reduced} in the sense that its leading mixed term is as low as possible (with respect to the term order $TP$) due to cancellation with leading mixed terms from previous polynomials.
      \item Reduce the ``updated'' $F[j,1]$ modulo the ideal $I$ to update $F[j,2]$ and $F[j,3]$.  
      \item If the new leading mixed term $F[j,2]$ is 0, we have a {global holomorphic function!}
      \end{itemize}
\item Determine which global holomorphic functions are ``new,'' so that the final list isn't redundant.
      \begin{itemize}
      \item Check that the new global holomorphic function $X_l$ is not in the ring $\CC[X_1,...,X_{l-1}]$
      generated by the previous functions.
      \item To do this we find a Groebner basis for the ideal $\langle X_1,...,X_l \rangle$ and compute partials
      to make sure we can't solve for the new function in terms of the previous ones.
      \end{itemize}
\item Find relations among the global holomorphic functions.  These will determine the (singular) 
geometry of the blow down.
\end{enumerate}


\subsection{A shortcut}\label{shortcut}

Of particular interest to us are the Intriligator--Wecht superpotentials \cite{wecht}.
As can be seen from the Table, each potential $W(x,y)$ has two possible expressions for $\partial_{w_1}E(\gamma,w_1)$, which
corresponds to exchanging $x \leftrightarrow y$ in the matrix model potential.
\begin{table}[h]
$$\begin{array}{c|c|c|c}
\mathrm{type} & W(x,y) & \partial_{w_1}E(\gamma,w_1) & \PP^1: (w_1,w_2), (v_1,v_2)\\
\hline
&&&\\
\Hat{O} & 0 & 0 & (x+\gamma y, 0), (\beta x + y, 0)\\
& & &\\
\Hat{A} & \12 y^2 & w_1 \leftrightarrow \gamma^2 w_1 & (x,0), (\beta x, x)\\
& & &\\
\Hat{D} & xy^2 & w_1^2 \leftrightarrow \gamma w_1^2 & (x,0), (\beta x, x^2)\\
& & &\\
\Hat{E} & \dfrac{1}{3} y^3 & \gamma^{-1}w_1^2 \leftrightarrow \gamma^2 w_1^2& (x,0), (\beta x, \beta x^2)\\
& & &\\
A_k & \dfrac{1}{k+1}x^{k+1}+\12 y^2 & \gamma^2 w_1^k + w_1 \leftrightarrow
\gamma^{1-k}w_1^k+\gamma^2w_1 & (0,0), (0,0)\\
& & &\\
D_{k+2} & \dfrac{1}{k+1}x^{k+1}+xy^2 & \gamma^2 w_1^k+w_1^2 \leftrightarrow
\gamma^{1-k} w_1^k + \gamma w_1^2 & (0,0), (0,0)\\
& & &\\
E_6 & \dfrac{1}{3}y^3+\dfrac{1}{4}x^4 & \gamma^{-1}w_1^2+\gamma^2 w_1^3 \leftrightarrow
\gamma^2 w_1^2 + \gamma^{-2} w_1^3 & (0,0), (0,0)\\
& & &\\
E_7 & \dfrac{1}{3}y^3+yx^3 & \gamma^{-1}w_1^2+\gamma w_1^3 \leftrightarrow
\gamma^2 w_1^2 + \gamma^{-1} w_1^3 & (0,0), (0,0)\\
& & &\\
E_8 & \dfrac{1}{3}y^3 + \dfrac{1}{5}x^5 & \gamma^{-1}w_1^2+\gamma^2 w_1^4
\leftrightarrow \gamma^2 w_1^2 + \gamma^{-3} w_1^4 & (0,0), (0,0)
\end{array}$$
\caption{Intriligator--Wecht superpotentials, and identification of corresponding resolved geometries}
\label{identify}
\end{table}

In all of the Intriligator--Wecht cases, we can find weights for the variables $\beta,v_1,v_2$ and $\gamma,w_1,w_2$ such that the transition functions
$$\beta=\gamma^{-1}, \skop v_1=\gamma^{-1}w_1, \skop v_2=\gamma^3 w_2+\partial_{w_1}E(\gamma,w_1),$$
are quasi-homogeneous.  For instance, in our above example (the $E_7$ case), the transition functions
$$\beta = \gamma^{-1}, \skop v_1 = \gamma^{-1} w_1, \skop v_2 = \gamma^3 w_2 + \gamma w_1^3 +
\gamma^{-1} w_1^2,$$
are quasi-homogeneous if we assign the weights
$$\begin{array}{c|c|c|c|c|c}
\beta & v_1 & v_2 & \gamma & w_1 & w_2 \\
1 & 3 & 5 & -1 & 2 & 8
\end{array}.$$
In particular, this means all elements of the ideal 
$$I = \langle \beta\gamma-1, v_1-\beta w_1, v_2-\gamma^3 w_2 - \gamma w_1^3 - \beta w_1^2 \rangle$$
are quasi-homogeneous in these weights, and all terms in the expression 
$$\beta^i v_1^j v_2^k \stackrel{\mathrm{mod\;}I}{\equiv} \pure(\beta,v_1,v_2) + 
\pure (\gamma,w_1,w_2) + \mixed, $$
will have the same weight.  

This immediately tells us that only combinations of monomials {\em of the same weight} can be used to cancel mixed
terms -- i.e. the global holomorphic functions we build will themselves be quasi-homogeneous.  This observation
cuts computational time immensely, since it means that in the {reduction routine} we need only cycle through lists
of polynomials of the same weight in order to reduce the order of the mixed terms.  In particular, we can run the algorithm in parallel for different weights, restricting ourselves to lists of monomials in $\CC[\beta,v_1,v_2]$ which are all in the same weighted degree.

For a detailed implementation (including actual Maple code) of the algorithm using this shortcut, see \cite{thesis}.

\section{Acknowledgments}
I would like to thank my advisor, David R. Morrison, for many helpful suggestions.
This work was supported by an NSF graduate fellowship and VIGRE.

\bibliographystyle{my-h-elsevier}

\end{document}